\documentclass[12pt]{elsart}
\usepackage{amsmath}
\usepackage{amssymb}
\usepackage{amsfonts}
\usepackage{graphics}
\usepackage{epsfig}
\usepackage{latexsym} 

\allowdisplaybreaks

\newcommand{\cal}{\mathcal}

 \newcommand{\Dlr}{\mbox{\parbox[b]{0cm}{$D$}\raisebox{1.7ex}
                       {${\,\scriptstyle{\leftrightarrow}}$}}}
 \newcommand{\Dl}{\mbox{\parbox[b]{0cm}{$D$}\raisebox{1.7ex}
                       {${\,\scriptstyle{\leftarrow}}$}}}
 \newcommand{\Dr}{\mbox{\parbox[b]{0cm}{$D$}\raisebox{1.7ex}
                       {${\,\scriptstyle{\rightarrow}}$}}}
 \newcommand{\Dlrsl}{\not{\hspace{-0.14cm}{\Dlr}}}  
 \newcommand{\pslash}{{\not{\hspace{-0.08cm}p}}}  
 \newcommand{\psib}{\bar{\psi}}
   
 \newcommand{\csw}{\, c_{\scriptscriptstyle{SW}}}  
 \newcommand{\ggcf}{\frac{g^2 C_F}{16 \, \pi^2}\; }  
   
 \newcommand{\Tterm}{\, T \left(p^2 /m^2 \right) }
 \newcommand{\minT}{\, \frac{m^2}{p^2} 
                    \left( 1 - T \left(p^2/m^2 \right)\right)}
   
 \newcommand{\mmpole}{\frac{m^2}{m^2 + p^2} }  
   
 \newcommand{\msb}{{\scriptscriptstyle{\overline{{\rm MS}}}}}  
 \newcommand{\MSB}{{\overline{{\rm MS}}}}  
 \newcommand{\cscl}{c_1^S}  
 \newcommand{\cmu}{c_1^V}  
 \newcommand{\cax}{c_1^A}  
 \newcommand{\ct}{\, c_1^H}  
 \newcommand{\pvstruc}{\frac{{\rm i}}{2}(\pslash\gamma_\mu\gamma_5 + 
                                     \gamma_\mu\gamma_5\pslash)}  
 \newcommand{\tenstruc}{\frac{{\rm i}}{2}(\pslash\sigma_{\mu\nu}\gamma_5 + 
                                     \sigma_{\mu\nu}\gamma_5\pslash)}  
 \newcommand{\mslog}{\, \ln \left( \frac{p^2 + m_\msb^2}{\mu^2} \right)}  
 \newcommand{\msTterm}{\, T \left(p^2 /m_\msb^2 \right) }
 \newcommand{\msminT}{\, \frac{m_\msb^2}{p^2}
                    \left( 1 - T \left(p^2/m_\msb^2 \right)\right)}
 \newcommand{\starminT}{\, \frac{m_\star^2}{M^2}
                    \left( 1 - T \left(M^2/m_\star^2 \right)\right)}
 \newcommand{\starTterm}{\, T \left(M^2 /m_\star^2 \right) }
 \newcommand{\starLapam}{\, L \left( a M , a m_\star \right) } 
 \newcommand{\calO}{ {\mathcal O} }
 \newcommand{\FT}{\;{\mathcal F}\;} 
 \newcommand{\ti}{ {\scriptscriptstyle{\rm TI}}} 
 \newcommand{\cswti}{\,{\csw^\ti}} 
 \newcommand{\cti}{\, c^\ti} 
 \newcommand{\ggcfti}{\frac{g_\ti^2}{16 \pi^2} C_F \,} 

\graphicspath{plots}
\begin{document}
\vspace{-1.0cm}
\begin{flushleft}
{\normalsize DESY 00-049} \hfill\\
{\normalsize TPR-00-07} \hfill\\
{\normalsize LU-ITP-2000/002} \hfill\\
{\normalsize MIT-CTP-2981} \hfill\\
{\normalsize HUB-EP-00/26} \hfill\\
{\normalsize July 2000} 
\end{flushleft}
\vspace{1.0cm}
\begin{frontmatter}
 \title{\bf Renormalisation and off-shell improvement
   in lattice perturbation theory}

  \author[MI,HH]{S. Capitani},
  \author[RG]{M. G\"ockeler},
  \author[HU]{R. Horsley},
  \author[LZ]{H. Perlt},
  \author[RG]{P. E. L.  Rakow}, 
  \author[HH,ZT]{G. Schierholz} and
  \author[LZ]{A. Schiller}
  
 \address[MI]{MIT, Center for Theoretical Physics, 
 Laboratory for Nuclear Science, \\
  77 Massachusetts Avenue, Cambridge, MA 02139, USA}
  \address[HH]{Deutsches Elektronen-Synchrotron DESY, D-22603
    Hamburg, Germany}
  \address[RG]{Institut f\"ur Theoretische Physik, Universit\"at
    Regensburg, D-93040 Regensburg, Germany} 
  \address[HU]{Institut f\"ur Physik, Humboldt-Universit\"at zu Berlin,
    D-10115 Berlin, Germany}
  \address[LZ]{Institut f\"ur Theoretische Physik, Universit\"at Leipzig,  
 D-04109 Leipzig, Germany}
  \address[ZT]{Deutsches Elektronen-Synchrotron DESY,\\
    John von Neumann-Institut f\"ur Computing NIC, D-15738
    Zeuthen, Germany} 

  \date{ }  

\begin{abstract}
 We discuss the improvement of flavour non-singlet point and one-link
 lattice quark operators, which describe the quark currents and the first
 moment of the DIS structure functions respectively.  
 Suitable bases of improved operators are given, and the corresponding  
 renormalisation factors and improvement coefficients
 are calculated in one-loop lattice perturbation theory, using the 
 Sheikholeslami-Wohlert (clover) action.
 To this order we achieve off-shell improvement by eliminating 
 the effect of contact terms.
 We use massive fermions, and our calculations are done keeping all terms 
 up to first order in the lattice spacing, for arbitrary $m^2/p^2$, in a 
 general covariant gauge. We also compare clover fermions 
 with fermions satisfying the Ginsparg-Wilson relation,
 and show how to remove $O(a)$ effects off-shell in this case too,
 and how this is in many aspects simpler than for clover fermions.
 Finally, tadpole improvement is also considered.
\end{abstract}
\end{frontmatter}

 \section{Introduction}  
   
 There has been considerable progress in obtaining realistic results  
 from numerical simulations in lattice QCD.
 A new generation of massively parallel computers promises  
 results that can be compared to a wide class of experimental  
 data. Nevertheless, the finiteness of the lattice spacing always leads to  
 systematic errors in the simulations. 
 Therefore there is great interest in improving lattice
 QCD calculations. A systematic improvement scheme, removing 
 discretisation errors order by order in the lattice spacing $a$, has
 been suggested by Symanzik~\cite{Symanzik}, and developed by L\"uscher
 and Weisz~\cite{LW} for on-shell quantities. 
   
   An $O(a)$ improved fermionic action which is widely 
 used in lattice Monte Carlo simulations is that proposed by 
 Sheikholeslami and Wohlert~\cite{SW}: 
 \begin{eqnarray} 
  S_{\rm imp}^{\rm MC} =  a^4 \sum_x \Big\{ && 
 \frac{1}{a} \, \psib(x) \psi(x) 
  - \frac{\kappa}{a} \sum_\mu \psib(x + a \hat{\mu}) \,
 U^\dagger_\mu(x) \left[1 + \gamma_\mu\right] \psi(x)
 \nonumber \\ &&
 - \frac{\kappa}{a} \sum_\mu \psib(x - a \hat{\mu}) \,
 U_\mu(x -a \hat{\mu}) \left[1 - \gamma_\mu\right] \psi(x)
 \nonumber \\ &&
  - 2 \kappa \,\csw \,g\,  \sum_{\mu\nu}\frac{a}{4}  
 \psib(x)\,\sigma_{\mu\nu} F_{\mu\nu}^{\rm clover}(x) \psi(x)
 \Big\} ,  
 \end{eqnarray}  
 where $\kappa$ is known as the hopping parameter and 
 $F_{\mu\nu}^{\rm clover} $ denotes the standard ``clover-leaf"
 form of the lattice field  strength.%
 \footnote{We will use the convention
 $\sigma_{\mu\nu}=({\rm i}/2)(\gamma_\mu \gamma_\nu- \gamma_\nu \gamma_\mu)$.}
 If the parameter $\csw$, which gives the strength of the higher-dimensional 
 operator, is correctly chosen this action has no $O(a)$
 errors for on-shell quantities such as hadron masses. 
 For perturbative calculations it is simpler to use a slightly different 
 normalisation: 
 \begin{eqnarray} 
  S_{\rm imp}^{\rm pert} =  a^4 \sum_x \Big\{ && 
 (m + m_c) \, \psib(x) \psi(x) 
 \nonumber \\ &&
  - \frac{1}{2 a} \sum_\mu \psib(x + a \hat{\mu}) \,
 U^\dagger_\mu(x) \left[1 + \gamma_\mu\right] \psi(x)
 \nonumber \\ &&
 - \frac{1}{2 a} \sum_\mu \psib(x - a \hat{\mu}) \,
 U_\mu(x -a \hat{\mu}) \left[1 - \gamma_\mu\right] \psi(x)
 \nonumber \\ &&
  - \csw \,g\,  \sum_{\mu\nu}\frac{a}{4}  
 \psib(x)\,\sigma_{\mu\nu} F_{\mu\nu}^{\rm clover}(x) \psi(x)
 \Big\} .  
 \label{SW_action} 
 \end{eqnarray}  
 The parameters of the two forms of the action are related by 
 \begin{eqnarray} 
 a m_c = \frac{1}{2 \kappa_c} \,, \label{kcdef} \\
 a m = \frac{1}{2 \kappa} - \frac{1}{2 \kappa_c} \,, 
 \end{eqnarray} 
 where $\kappa_c$ is the critical value of the hopping parameter,
 at which chiral symmetry is approximately restored. 

   Simply improving the action does not remove $O(a)$ errors from 
 operator matrix elements. To do this the operators must also 
 be improved by adding higher dimensional irrelevant 
 operators with appropriate
 improvement coefficients. The operators also need to be renormalised. 
 In this paper we will discuss the perturbative renormalisation and 
 improvement of bilinear quark operators.

   However, the action~(\ref{SW_action}) with its single tunable
  improvement parameter $\csw$ only improves on-shell 
 quantities. Off-shell quantities still have $O(a)$ errors, which 
 arise from short-distance ``contact" terms.  We will
 show how the contact terms can be removed at the one-loop level
 of lattice perturbation theory, and off-shell quantities
 free of $O(a)$ discretisation errors can be extracted from    
 Green's functions. 

   There are several reasons why it would be desirable to 
 understand the improvement of off-shell quantities. In particular
 the non-perturbative renormalisation suggested in~\cite{mart1}  
 involves comparing lattice measurements of off-shell Green's
 functions with continuum perturbation theory results~\cite{nonpert} 
 in order to relate lattice quantities to conventional
 renormalisation schemes such
 as $\MSB$. This matching will work best at large virtualities, 
 where the running coupling constant is small, and the effects of 
 non-perturbative phenomena such as chiral symmetry breaking have
 died away. It is obviously desirable to remove the discretisation
 errors in the off-shell lattice Green's functions before making the 
 comparison with the continuum. 
  Even within perturbation theory it is easier to 
 calculate Green's functions at $p^2 \gg m^2$ than in the 
 region where $p^2$ and $m^2$ are comparable. 

   Our strategy is to look at the tree-level results for the Green's 
 functions, and see what $O(a)$ effects are present, and what has to
 be done to remove them. We then look at the one-loop perturbative
 results, and see whether the tree-level procedure still works. We 
 find that at one particular value of the clover coupling the $O(a)$
 effects are of the same form as in tree-level, and that then we 
 can remove $O(a)$ effects completely, and find improved Green's 
 functions that are free of $O(a)$ discretisation errors, both 
 on-shell and off-shell. 

   Our aim is to find perturbative expressions at one-loop
 for the  $\MSB$-scheme renormalisation factors  
 and for the improvement coefficients. 
  To do this we have to compute each Feynman
 diagram including all $O(a)$ terms. These results are applicable to
 both quenched and dynamical calculations of flavour non-singlet 
 matrix elements.

 In this paper we consider the complete set of point operators
 \begin{equation}  
 \psib (x) \Gamma_i \psi (x),
 \label{locop}  
 \end{equation}
 with 
 \begin{equation}
 \Gamma_i=1,\;\gamma_5,\; \gamma_\mu,\;   
 \gamma_\mu\gamma_5,  
 \; \sigma_{\mu\nu}\gamma_5.  
 \end{equation}  
 For one-link operators we discuss the physically interesting 
 case of the leading-twist operators
 occurring in the operator product expansion for the moments of the 
 hadronic structure functions~\cite{qcdsf1}.  
  We consider the operators which measure the first moment
 of the unpolarised and polarised structure functions: 
 \begin{equation}  
 \frac{1}{2} \psib\gamma_\mu \Dlr_\nu \psi, \qquad  
 \frac{1}{2} \psib\gamma_\mu \gamma_5 \Dlr_\nu \psi ,  
 \label{nonlocop}  
 \end{equation}  
 where symmetrisation over $\mu$ and $\nu$ and removal of trace terms
 is always to be understood.

 The perturbative 
 renormalisation of improved point operators has been discussed by
 several groups~\cite{heatli,gabr,borelli}. They use the tree-level
 values for the operator improvement coefficients, $c_i$, (defined 
 below) and for    
 the coefficient $\csw$ in the  Sheikholeslami-Wohlert action.
  The same settings have  been used to   
 calculate the renormalisation factors for the one-link~\cite{capi1}
 and two-link~\cite{capi2} quark operators in the chiral limit, 
 performing on quark operators the transformation discussed
 in~\cite{borelli}. In this  
 paper we  present  the $Z$ factors with coefficients  $c_i$ and
 $\csw$ kept arbitrary. This allows us to determine the perturbative  
 contributions of the various terms and their relative magnitudes. Moreover,  
 this will enable us to implement tadpole improved perturbation theory.  
      
 This paper is organised as follows. 
 In Sect.~2 we give the operator bases for the improvement of the 
 point and one-link operators. 
 In Sect.~3 we present a method with which to improve the lattice quark 
 propagator off-shell by taking care of contact terms, and 
 in Sect.~4 we extend this procedure to improve off-shell quark bilinear 
 operators as well.
 In Sect.~5 we compare with fermions satisfying the Ginsparg-Wilson
 relation,
 and show how to remove $O(a)$ effects off-shell in this case too.
 Finally, in Sect.~6 we apply tadpole improvement to our perturbative
 results, and in Sect.~7 we present our conclusions. 
 The (sometimes cumbersome) complete results for renormalisation factors
 and improvement coefficients are collected in the Appendix. 

 \section{Bases for improved operators \label{basis} }  
   
 In this section we write down a general operator  
 basis for the improvement of quark operators. The base operators must
 have the same symmetry properties as the unimproved ones, i.e. their  
 transformations under the hypercubic group and charge conjugation
 are determined by the original operator.  
   
 First we consider the five point operators of eq.~(\ref{locop}). 
 Subject to the symmetry constraints   
 we find the following improved operators:  
 \begin{eqnarray}
 \left( \calO^S \right)^{\rm imp} &=& 
 \left(\psib\psi\right)^{\rm imp} = 
 (1 + a\,m\,c_0)  \psib\psi -   
 \frac{1}{2} a c_1 \psib\Dlrsl\psi \,,  \label{opdef} \\  
 \left( \calO^P \right)^{\rm imp} &=&
 \left(\psib\gamma_5\psi\right)^{\rm imp}  
 = (1 + a\,m\,c_0)  \psib\gamma_5\psi +   
 \frac{1}{2}a c_2 \partial_\mu \Big(\psib\gamma_\mu\gamma_5\psi\Big)
 \,, \\
 \left( \calO^V_\mu \right)^{\rm imp} &=&  
 \left(\psib\gamma_\mu\psi\right)^{\rm imp}  
 = (1 + a\,m\,c_0) \psib\gamma_\mu\psi -   
 \frac{1}{2} a c_1 \psib\Dlr_\mu \psi \nonumber \\
 && +\frac{1}{2} a {\rm i} c_2 \partial_\lambda\Big(\psib
 \sigma_{\mu\lambda}\psi\Big) \,, \\
 \left( \calO^A_\mu \right)^{\rm imp} &=&
 \left(\psib\gamma_\mu\gamma_5\psi\right)^{\rm imp} =   
   (1 + a\,m\,c_0) 
 \psib\gamma_\mu\gamma_5\psi \nonumber \\
 && -\frac{1}{2} a {\rm i} c_1 \psib\sigma_{\mu\lambda}
 \gamma_5\Dlr_\lambda \psi +  
 \frac{1}{2} a c_2 \partial_\mu\Big(\psib\gamma_5\psi\Big)\,, \\ 
 \left( \calO^H_{\mu\nu} \right)^{\rm imp} &=&
 \left(\psib\sigma_{\mu\nu}\gamma_5\psi\right)^{\rm imp}=  
  (1 + a\,m\,c_0) 
  \psib\sigma_{\mu\nu}\gamma_5\psi \nonumber \\
 && +\frac{1}{2} a {\rm i} c_1\psib  
      \left(\gamma_\mu\Dlr_\nu -\gamma_\nu\Dlr_\mu\right)\gamma_5 \psi  
 + \frac{1}{2} a {\rm i} c_2 \, \epsilon_{\mu\nu\lambda\tau} \, 
 \partial_\tau\Big(\psib\gamma_\lambda\psi\Big) ,
 \end{eqnarray}  
 where $m$ is the bare fermion mass and $\Dlr \equiv \Dr - \Dl$ is the
 symmetric covariant derivative. We have used the lattice definitions
 \begin{eqnarray}
  \Dr_\mu \psi (x)  &=& \frac{1}{2a} \Big[ U_\mu (x) \psi (x + a \hat{\mu}) 
 - U_\mu^\dagger (x - a \hat{\mu}) \psi (x - a \hat{\mu}) \Big] \,,\\
 \psib (x) \Dl_\mu &=& \frac{1}{2a} \Big[ 
    \psib (x + a \hat{\mu}) U_\mu^\dagger (x)
  - \psib (x - a \hat{\mu}) U_\mu (x - a \hat{\mu}) \Big] \,. \nonumber 
 \end{eqnarray}
 The $c_2$ terms in the above equations are irrelevant for 
 forward matrix elements, which are all that we consider. 
 Therefore we are left with the expressions in Table~\ref{basistab}. 

\renewcommand{\arraystretch}{2}

\begin{table}[Htb]
\begin{center}
\begin{tabular}{|c||l|}
\hline
 & Improvement basis \\ \hline 
 $S$ & $ (1 + a\,m\,c_0)  \psib\psi -   
 \frac{1}{2} a c_1 \psib\Dlrsl\psi $ \\
 $P$  & $ (1 + a\,m\,c_0)  \psib\gamma_5\psi $ \\
 $V$  & $ (1 + a\,m\,c_0) \psib\gamma_\mu\psi -   
 \frac{1}{2} a c_1 \psib\Dlr_\mu \psi $ \\
 $A$  & $ (1 + a\,m\,c_0) \psib\gamma_\mu\gamma_5\psi 
 -\frac{1}{2} a {\rm i} c_1 \psib\sigma_{\mu\lambda}
 \gamma_5\Dlr_\lambda \psi$ \\ 
 $H$  & $ (1 + a\,m\,c_0) \psib\sigma_{\mu\nu}\gamma_5\psi 
 +\frac{1}{2} a {\rm i} c_1\psib  
 \left(\gamma_\mu\Dlr_\nu -\gamma_\nu\Dlr_\mu\right)\gamma_5 \psi $ \\
 \hline 
\end{tabular} 
\end{center}
\vspace{0.5 cm}
\caption{The improvement bases for the scalar (S), pseudoscalar (P),
 vector (V), axial (A) and tensor (H) operators.}\label{basistab}
\vspace{0.5 cm}
\end{table} 

 We include the terms proportional 
 to $c_0$ so that we can get $m$-independent renormalisation 
 constants and at the same time maintain $O(a)$ improvement.
  
 Using the scalar operator as an example, there is an equation of
 motion that says that for on-shell measurements 
 $ \psib\Dlrsl\psi + \eta m  \psib\psi = 0 $, 
 where $\eta$ is a coefficient that depends on $g$.
 So we can compensate for changes in $c_1$ by making 
 changes in $c_0$, which would allow us to eliminate one of the 
 improvement terms if we were only interested in on-shell quantities. 
 This equation of motion means that $c_0$ is linear in $c_1$ if we 
 parameterise our operators as in eq.~(\ref{opdef}). If we use 
 other parameterisations, for example
 $(1 + a b  m) ( \psib\psi 
 - \frac{1}{2} a c_1' \psib\Dlrsl\psi )$~\cite{chiral},
 we would no longer find that $b$ was linear in $c_1'$.
 
 We also consider the conserved vector current: 
 \begin{eqnarray}
 (J_\mu)^{\rm imp}  &=& \frac{1}{2}  
 \psib(x) (\gamma_\mu -1)  U_\mu(x) \psi(x+a {\hat \mu})
 + \frac{1}{2} 
   \psib(x+a {\hat \mu}) (\gamma_\mu +1) U^\dagger_\mu(x) \psi(x)
 \nonumber \\   
 && +\half a {\rm i} c_2 \partial_\lambda\Big(\psib
 \sigma_{\mu\lambda}\psi\Big).
 \end{eqnarray}
  We know that this should need no improvement for 
 forward matrix elements, because the only improvement term, $c_2$,
 is the coefficient of a total derivative, and so has no effect on 
 forward matrix elements. Thus $J_\mu$
 provides a useful check of our improvement method.

 Next we consider the one-link operators of eq.~(\ref{nonlocop}).
 Here we choose as a basis for the improved unpolarised operator
 \begin{eqnarray}  
 \left( \calO_{\mu\nu} \right)^{\rm imp } & = &  
 \left(1+ a\, m\, c_0 \right) \frac{1}{2}
 \psib\gamma_\mu \Dlr_\nu \psi +   
 \frac{1}{8}\,a\,{\rm i} c_1  \sum_\lambda \psib 
 \sigma_{\mu\lambda}\left[\Dlr_\nu,\Dlr_\lambda\right]\psi \nonumber \\  
 & & -\frac{1}{8} \,a \,c_2  
  \psib\left\{\Dlr_\mu,\Dlr_\nu\right\}\psi   +   
  \frac{1}{4}\,a \,{\rm i}\,c_3 \sum_\lambda \partial_\lambda 
 \Big(\psib\sigma_{\mu\lambda}\Dlr_\nu\psi\Big) . \label{imp2}   
 \end{eqnarray}  
 This operator basis is the same for the two possible irreducible 
 representations of the lattice hypercubic group to which the original
 operator may belong:  $\tau_3^{(6)}$ (non-diagonal, $\mu \ne \nu$)
 and  $\tau_1^{(3)}$ (diagonal, $\mu = \nu$).
 (Our notation for the irreducible representations of the
 lattice hypercubic group follows~\cite{hypercube}.)  
 In the case of the polarised structure function 
 we find that the improvement terms allowed by the hypercubic symmetry 
 are different for the representations $\tau_4^{(6)}$ 
 (non-diagonal, $\mu \ne \nu$)
 and $\tau_4^{(3)}$ (diagonal, $\mu = \nu$).
 When $\mu \ne \nu$ the improved operator has the form
 \begin{eqnarray} 
 \left( \calO_{\mu\nu}^5 \right)^{\rm imp} & = &  
 \left(1+ a\, m\, c_0 \right) \frac{1}{2} 
 \psib\gamma_\mu\gamma_5 \Dlr_\nu \psi     
 - \frac{1}{4}\, {\rm i} a\, c_1\, \psib\sigma_{\mu \nu}  
 \gamma_5 \Dlr_\nu^2 \psi \label{imp5} \\   
 & & 
 -\frac{1}{8}\,a \,{\rm i}\,c_2 \, \sum_{\lambda \ne \mu, \nu} 
 \psib\sigma_{\mu\lambda} 
  \gamma_5\left\{\Dlr_\lambda,\Dlr_\nu\right\}\psi      
 + \frac{1}{4}\,a \, c_3 \partial_\mu\Big(\psib 
 \gamma_5\Dlr_\nu\psi\Big) \nonumber ,
 \end{eqnarray}
 whereas in the traceless diagonal case one has only one
 improvement term in the forward case, so there is no $c_2$, and 
 thus
 \begin{eqnarray}  
 \left( \calO_{\mu\mu}^5 \right)^{\rm imp} & = &  
 \left(1+ a\, m\, c_0 \right)  \frac{1}{2}
 \psib\gamma_\mu\gamma_5 \Dlr_\mu \psi 
 -\frac{1}{8}\,a \,{\rm i}\,c_1 \, \sum_\lambda \psib\sigma_{\mu\lambda} 
  \gamma_5\left\{\Dlr_\lambda,\Dlr_\mu\right\}\psi      
 \nonumber\\ & & 
 + \frac{1}{4}\,a \, c_3 \partial_\mu\Big(\psib 
 \gamma_5\Dlr_\mu\psi\Big)\label{imp4} .
 \end{eqnarray}  
 Here repeated $\mu$ and $\nu$
 indices are not summed over. We will 
 always construct a traceless operator when the indices are equal by 
 using the combination
 $\frac{1}{2} \psib \frac{1}{2} ( \gamma_\mu \gamma_5 \Dlr_\mu 
 -\gamma_\nu \gamma_5 \Dlr_\nu )\psi$,
 and a similar one in the unpolarised case. 

The coefficients $c_0, \ldots, c_2$ can be appropriately 
determined using the method explained in the following sections 
so that the desired improvement is achieved.

 \section{Improving the quark lattice propagator}  

 \subsection{Method \label{propmethod}} 
   
  Even when the fermion action has been improved for on-shell 
 quantities there are still $O(a)$ effects present in off-shell 
 quantities such as the fermion propagator at a general Euclidean 
 momentum $p$.  
 In this section we will discuss how to find an improved fermion 
 propagator off-shell. First we will look at the tree-level   
 Wilson propagator and show how to remove its off-shell  
 discretisation errors, then we will generalise this improvement 
 method to the interacting case.  
  
   We are used to writing down expressions for $S^{-1}$, 
 the inverse quark propagator. In   $S^{-1}$  
 the main $O(a)$ effect is the addition of the momentum-dependent 
 Wilson mass term. However it is also instructive to look at the quark 
 propagator $S$ itself, rather than the inverse propagator.  
 
   Let us start by looking at the propagator at tree-level. From 
 the expression for the inverse propagator
 \begin{eqnarray} 
  (S^{{\rm tree}})^{-1}(p,m) &=&  
 \frac{{\rm i}}{a} \sum_{\mu} \gamma_{\mu} \sin(a p_\mu)  + m  
 + \frac{1}{a} \sum_{\mu} \left( 1 - \cos (a p_\mu) \right) \nonumber \\  
 &=& {\rm i} \pslash + m + \half a  p^2  + O(a^2) \\ 
  &=& ({\rm i}\pslash + m - \half a m^2 ) (1 + a  m )  
 (1 - \half a  ({\rm i}\pslash + m ) ) + O(a^2) \nonumber   
 \end{eqnarray} 
  we derive that
 \begin{eqnarray} 
 S^{{\rm tree}}(p,m) &=& 
 \frac{1 - a   m}{{\rm i}\pslash + m - \half a  m^2 }  + \frac{a}{2}  
 + O(a^2) \nonumber \\ 
 &\equiv& (1 - a m) S_\star^{{\rm tree}}(p, m_\star) + \frac{a}{2} 
 + O(a^2).  
 \label{treemomprop}  
 \end{eqnarray} 
 The tree-level lattice propagator consists then of two parts, one part 
 is proportional to a normal continuum propagator with a mass $m_\star \equiv  
 m - \half a m^2$, and the other part is a momentum independent term. 
 The nature of these two parts becomes even more clear when we write them 
 in position space:\footnote{On the lattice we define 
 $\delta (x-y) \equiv \delta_{x,y} /a^4 $, where $ \delta_{x,y}$ is 
 the Kronecker delta function.}  
 \begin{equation} 
 S^{{\rm tree}}(x,y,m) = 
 (1 - a  m) S_\star^{{\rm tree}}(x,y,m_\star) 
  + \frac{1}{2} a \delta(x- y)  + O(a^2) .  
 \label{treexprop}  
 \end{equation}  
 We see that (except at short distances, where an additional contact 
 term appears) the lattice Wilson-fermion propagator is proportional to  
 $S_\star$, which has the form of a continuum propagator 
 with an ``improved" mass $m_\star$, and which has no $O(a)$
 discretisation errors. We will always use $\star$ to mark 
 bare quantities which have been $O(a)$ improved.

 This concentration of the $O(a)$ effects at short distance is what we 
 should expect, in fact the fermion propagator at $|x-y| \gg a$ is an 
 on-shell quantity, so it should be automatically improved when the action 
 is improved. It is only at short distances of order $a$ that the lattice  
 propagator has a different form from the continuum propagator.  
 The necessity of subtracting a $\delta$ function from the lattice 
 propagator to obtain an improved propagator has been discussed 
 in~\cite{mart1}.  
 
   What should we expect beyond tree level? Let us write the inverse 
 fermion propagator as a series in the lattice spacing $a$:  
\begin{equation} 
 S^{-1}(p,m) = \sigma_0(p,m) + a \sigma_1(p,m) + O(a^2) ,
 \label{sig01def}  
\end{equation}  
 where the coefficients $\sigma_0$ and $\sigma_1$  
 are power series in $g^2$. On-shell improvement tells us that  
 the lattice fermion propagator should be proportional to a continuum  
 fermion propagator except at short distances, so we expect equations  
 of the same form as eqs.~(\ref{treemomprop}) and (\ref{treexprop}) to hold,  
 though only at the value of $\csw$ corresponding to on-shell  
 improvement. Thus the propagator should satisfy 
 \begin{eqnarray} 
 S(p,m) &=& \frac{1 + a b_\psi m}{\sigma_0(p,m_\star)}  
 + \frac{a}{2} \lambda_\psi + O(a^2) \nonumber \\ 
 &\equiv& (1 + a b_\psi m ) S_\star(p,m_\star)  
 + \frac{a}{2} \lambda_\psi + O(a^2) ,  
 \label{simprove}  
 \end{eqnarray}  
 where the improved bare mass $m_\star$ is related to $m$ through 
 \begin{equation} 
 m_\star = m (1 + a b_m m).   
 \label{mstar}  
 \end{equation}  
 The improvement coefficients $b_\psi, b_m$ and $\lambda_\psi$ are 
 independent of $p$ and $m$, and should only depend on the coupling 
 constant $g^2$. By comparing with eq.~(\ref{treemomprop}) we see
 that the tree-level values are  
 \begin{equation}
 b_\psi =-1, \qquad b_m=-1/2, \qquad \lambda_\psi=1 .
 \end{equation} 
 The propagator $S_\star$ is free of $O(a)$ effects so we call it 
 the improved fermion propagator. Later, when we come to define 
 renormalisation constants, we will 
 always define them in terms of the improved bare propagator $S_\star$ 
 and improved bare mass $m_\star$.  
 
 Taking the inverse of eq.~(\ref{simprove}) gives
 \begin{eqnarray} 
 S^{-1}(p,m) &=&  
 \sigma_0(p,m_\star) \left( 1 - a b_\psi m \right)  
 \left( 1 - \half a \lambda_\psi \sigma_0(p,m_\star)  \right)   
 + O(a^2) \label{inverse_improve} \\ 
 &=&   
 \left( \sigma_0(p,m) +  
 a b_m m^2 \frac{\partial}{\partial m} \sigma_0(p,m) \right) \nonumber \\
 && \times \left( 1 - a b_\psi m \right)  
 \left( 1 - \half a \lambda_\psi \sigma_0(p,m)  \right)   
 + O(a^2) ,\nonumber  
\end{eqnarray} 
 so that dropping terms of order $a^2$ we get
\begin{eqnarray} 
 S^{-1}(p,m)  =   \sigma_0(p,m) &+& a \Big[  
 - \half \lambda_\psi \left( \sigma_0(p,m)\right)^2  
 -  b_\psi m  \sigma_0(p,m) \\ 
 && + b_m m^2 \frac{\partial}{\partial m} \sigma_0(p,m) \Big] 
 + O(a^2)  \nonumber .
 \end{eqnarray} 
 Comparing with eq.~(\ref{sig01def}) we see that the improvement  
 prescription~(\ref{simprove}) can only work if the non-linear relation  
 \begin{equation}  
 \sigma_1(p,m) =  
 - \half \lambda_\psi \left( \sigma_0(p,m)\right)^2 
 - b_\psi m  \sigma_0(p,m) 
 + b_m m^2 \frac{\partial}{\partial m} \sigma_0(p,m)  
 \label{sig1eqn}  
 \end{equation}  
 is satisfied. In subsection \ref{prop1loop} we shall see that
 this is indeed the case in one-loop perturbation theory. 

 The pole mass of the fermion is the $p$ value  
 where $\sigma_0(p,m_\star)$ vanishes, so for an on-shell fermion 
 the factor $( 1 - \half a \lambda_\psi \sigma_0(p,m_\star) )$  
 simply reduces to 1. Therefore
 we can see from eq.~(\ref{inverse_improve}) that the improvement  
 coefficient $\lambda_\psi$ only has an effect when the fermion 
 is off-shell, so we only need to know $ \lambda_\psi$  
 if we are interested in extracting numbers from off-shell lattice 
 measurements.   

 From eq.~(\ref{simprove}) we can find an explicit expression for 
 $S_\star$:  
 \begin{equation}
    S_\star(p,m_\star) = \frac{1}{1 + a m b_\psi  }
    \left( S(p, m) -\frac{a}{2} \lambda_\psi \right)
 = \frac{1}{\sigma_0(p,m_\star)} 
  \label{imp0_prop}
 \end{equation}
 and 
 \begin{eqnarray}
 S^{-1}_\star(p, m_\star) &=& (1 + a m b_\psi) \left( S^{-1}(p, m)
 +  S^{-1}(p, m) \frac{a}{2} \lambda_\psi S^{-1}(p, m)
 \right) \nonumber \\
 &=& \sigma_0(p, m_\star). 
 \label{imp_inv} 
 \end{eqnarray}
 The reason we are interested in 
  $S_\star (p,m_\star)$ is that it is a quantity
 free of $O(a)$ effects which can be constructed from the quark
 propagator $S (p,m)$, and the latter is something we can measure
 from non-perturbative simulations on the lattice. 
 We will find that eq.~(\ref{sig1eqn}) is satisfied at one particular
 value of the clover coefficient $\csw$. At this $\csw$ value one can 
 use eq.~(\ref{imp0_prop}) or equivalently eq.~(\ref{imp_inv}) to extract 
 $\sigma_0$ from lattice measurements. The clover action does not have 
 enough tunable parameters to make the off-shell fermion propagator 
 free of $O(a)$ effects, but this does not really matter, because 
 equations such as eq.~(\ref{imp_inv}) never-the-less allow us to 
 recover the improved off-shell propagator from quantities which 
 we can measure.  

 Up till now, we have only discussed the improvement of the fermion
 propagator. The propagator and mass still have to be renormalised. 
 The renormalised improved quark mass and propagator are given by
 \begin{eqnarray}
   m_R(\mu^2) &=&  Z_m(\mu^2) \; m_\star 
  = Z_m(\mu^2) \; m (1 + a m b_m  ) ,
 \label{renimp_mass} \\
    S_R(p,m_R;\mu^2) &=& \frac{ S_\star(p, m_\star)}{Z_2(\mu^2)} 
 = \frac{1}{Z_2(\mu^2) (1 + a m b_\psi ) }
    \left( S (p, m) -\frac{a}{2} \lambda_\psi \right) . 
  \label{renimp_prop} 
 \end{eqnarray} 
    
 \subsection{One-loop results for the quark propagator   
 \label{prop1loop}} 
 
 We now want to see if the propagator  
 improvement scheme suggested in eq.~(\ref{simprove}) holds in 
 one-loop perturbation theory, and  
 to calculate the improvement coefficients and renormalisation 
 factors to $O(g^2)$.  

   Our calculations are carried out in a general covariant gauge, 
 where the gluon propagator is
 \begin{equation}
 G(k) = \frac{ \delta_{\mu \nu}}{{\hat k}^2} 
 - (1 - \alpha) \frac{ {\hat k}_\mu {\hat k}_\nu} {({\hat k}^2)^2} , 
 \end{equation} 
 with ${\hat k}_\mu \equiv \frac{2}{a} \sin (a k_\mu/2)$. 
 The Feynman gauge corresponds to $\alpha = 1$, the Landau gauge 
 to $\alpha = 0$. 

  We can write  the inverse propagator in the form
 \begin{equation}  
 S^{-1}(p,m) =  m + {\rm i} \pslash + \frac{a}{2} p^2  
 - \ggcf \Big( \rho_{0}(p,m) 
 + a \rho_1(p,m) \Big) + O(a^2,g^4) ,     
 \label{rhodef} 
 \end{equation}  
 where $C_F = (N_c^2 -1)/(2 N_c)$ for gauge group $SU(N_c)$ . 
 Comparing eq.~(\ref{rhodef}) with eq.~(\ref{sig01def}) we see that 
 \begin{eqnarray} 
 \sigma_0(p,m) &=& {\rm i} \pslash + m   
 - \ggcf \rho_0(p,m)  +O(g^4) , \nonumber \\ 
 \sigma_1(p,m) &=& \frac{1}{2} p^2    
 - \ggcf \rho_1(p,m)  +O(g^4).   
 \label{rhoval}  
 \end{eqnarray} 
 We also expand the improvement coefficients to first order in  $g^2$:  
 \begin{eqnarray}  
\lambda_\psi &=& 1 + \ggcf d_{\lambda_\psi}   
 + O(g^4) , \nonumber \\  
 b_\psi &=& -\left[ 1 + \ggcf d_{b_\psi}  
       + O(g^4) \right] , \nonumber \\  
 b_m &=& - \frac{1}{2} \left[ 1 + \ggcf d_{b_m}  
         + O(g^4) \right].    
\label{ddef}  
 \end{eqnarray}  
 
 If we now substitute (\ref{rhoval}) and (\ref{ddef}) into the quadratic 
 equation (\ref{sig1eqn}), we find that the $\rho$ functions must obey
 the following linear condition if the improvement procedure suggested 
 in eq.~(\ref{simprove}) works:  
 \begin{eqnarray}  
 \rho_1(p,m) &+& {\rm i} \pslash \rho_0(p,m) + 
 \frac{m^2}{2} \frac{\partial}{\partial m} \rho_0(p,m) =  \label{rho1eqn} \\  
 & & -d_{\lambda_\psi} \frac{p^2}{2}  
 +\left( d_{\lambda_\psi} -d_{b_\psi} \right) {\rm i} \, \pslash m  
 +\left( d_{\lambda_\psi} -2 d_{b_\psi} + d_{b_m} \right) \frac{m^2}{2}
 \nonumber .
 \end{eqnarray}  
  
 The explicit expressions for $\rho_0$ and $\rho_1$ can be read from
 the one-loop expression for the fermion propagator up to $O(a)$:
 \begin{eqnarray}
 \lefteqn{ 
 S^{-1}(p,m)  = {\rm i} \, \pslash + m + \frac{a}{2} p^2 }\nonumber \\
  &-& {\rm i}\,\pslash \, \ggcf \Big[ \,
      16.64441 - \alpha   - 2.24887\csw - 1.39727\csw^2
   + \alpha \,L(a p, a m)
  \nonumber \\
 & &   +  \alpha \frac{m^2}{p^2} \left( 1 - \Tterm \right)
      \Big]  \nonumber \\
      & - &  m  \ggcf \Big[ \,
     11.06803 - 2\,\alpha
   - 9.98679\csw - 0.01689\,{ \csw^2}
       \nonumber \\
 & &  + ( 3 + \alpha  ) \,L(a p, a m) +  ( 3 + \alpha ) \Tterm
  \Big]
 \nonumber \\
  &-& a\, p^2\, \ggcf \Big[
     7.13891 - 0.07187\,\alpha  + 0.48567\csw - 0.08173\csw^2
 \nonumber \\ &&
 - \half \left( 3 - 2 \,\alpha  -  3 \csw \right) \, L(a p, a m)
        \Big] \nonumber \\ 
 &-& {\rm i} \, a\,m \pslash\,\ggcf \Big[
   -6.34664 + 0.14375\,\alpha  - 1.48503\csw + 1.28605\csw^2
 \nonumber \\ &&
 - \half \left( 3 + 2 \, \alpha + 3 \csw \right) \, L(a p, a m)
          - ( \alpha + 3\csw ) \Tterm
\nonumber \\ & &  
 - \half \left( 3 + 4 \,\alpha - 3 \csw \right)
    \frac{m^2}{p^2}\, \left( 1 - \Tterm \right)
 \Big] \label{Sinv1loop} \\ 
  &-& a m^2\,\ggcf \Big[ -14.03413 + 1.07187\,\alpha
   + 15.48574\csw - 1.52344\csw^2
        \nonumber \\ & &
 - \half \left( 9 + \alpha -  6 \, \csw \right) L(a p, a m)
 - \half \left( 12 + 5\,\alpha - 3 \csw \right) \Tterm  \Big] . \nonumber
 \end{eqnarray}
 They are
 \begin{eqnarray}
\rho_0(p,m)  =
  & {\rm i}\,\pslash & \,\Big[ \,
      16.64441 - \alpha   - 2.24887\csw - 1.39727\csw^2 \nonumber \\
 & &    + \alpha \,L(a p, a m) 
     + \alpha \frac{m^2}{p^2} \left( 1 - \Tterm \right)
      \Big]  \nonumber \\
      & + \, m  & \Big[ \, 11.06803 - 2\,\alpha
   - 9.98679\csw - 0.01689\,{ \csw^2}
     \nonumber \\
 & & + ( 3 + \alpha  ) \,L(a p, a m) +  ( 3 + \alpha ) \Tterm \Big]
\label{r0ans}
 \end{eqnarray}
and
 \begin{eqnarray}
  \rho_1(p,m)
   &=&   p^2  \Big[
  7.13891 - 0.07187\,\alpha  + 0.48567\csw - 0.08173\csw^2 \nonumber \\
 && - \half \left( 3 - 2\,\alpha  - 3 \csw \right) \, L(a p, a m)
        \Big]  \nonumber \\ 
 & + {\rm i}\,m \pslash & \Big[
   -6.34664 + 0.14375\,\alpha  - 1.48503\csw + 1.28605\csw^2 \nonumber \\
 && - \half \left( 3 + 2\,\alpha  +  3 \csw \right) \, L(a p, a m)
 - ( \alpha + 3\csw ) \Tterm \nonumber \\ 
 && - \half \left( 3 + 4\, \alpha - 3 \csw \right) 
    \frac{m^2}{p^2}\, \left( 1 - \Tterm \right)
 \Big] \nonumber \\     
 & + m^2 &\Big[ -14.03413 + 1.07187\,\alpha   
   + 15.48574\csw - 1.52344\csw^2 \nonumber \\ 
 && - \half \left( 9 + \alpha -  6\csw \right) L(a p, a m) \nonumber \\
 && - \half \left( 12 + 5 \,\alpha - 3 \csw \right) \Tterm  \Big] ,
 \label{r1ans}
 \end{eqnarray}
 where 
 \begin{eqnarray}
  T(x) &\equiv& \ln(1+x) / x \,, \nonumber \\
  L(x,y) &\equiv& \gamma_E - F_0 + \ln (x^2 + y^2) 
 \label{TLdef} 
 \end{eqnarray} 
  with $  F_0 = 4.369225\cdots$ and $\gamma_E = 0.577216\cdots$.
 Previously~\cite{caplat97} we calculated the fermion propagator
 in the limit $m^2 \ll p^2$, but eqs.~(\ref{r0ans}) and (\ref{r1ans}) 
 are valid for any ratio $m^2/p^2$ (but $a^2 m^2$ and $a^2 p^2$ must
 both be small).  

 Despite the complicated form of eqs.~(\ref{r0ans}) and (\ref{r1ans})  
 it can  be checked that at $\csw = 1$, and only at this value, 
 eq.~(\ref{rho1eqn}) is satisfied by $\rho_0$ and  
 $\rho_1$, and hence eq.~(\ref{sig1eqn}) is fulfilled. 
 This allows us to fix the improvement coefficients, which   
 in a general covariant gauge have the values
 \begin{eqnarray} 
 \csw &=& 1 + O(g^2) , \nonumber \\   
 \lambda_\psi &=&  1 +  \ggcf (10.91085 - 1.85625 \; \alpha)  
 + O(g^4) , \nonumber \\  
 b_\psi &=& - \left[ 1 +  \ggcf 16.39210  
 + O(g^4) \right] , \nonumber \\  
 b_m &=& -\frac{1}{2} \left[ 1 +  \ggcf 22.79406 + O(g^4) \right].   
 \label{propcoef} 
 \end{eqnarray} 
 Both $b$ coefficients are gauge invariant.\footnote{There was a  
 mistake in $b_m$ in~\cite{caplat97}, which has been corrected here.}
 
 In addition to the propagator, 
 eq.~(\ref{Sinv1loop}), our calculation also gives us a
 one-loop expression for the critical coupling~\cite{qcdsf3,qcdsf3b} 
 \begin{equation}  
  \kappa_c = \frac{1}{8} \left[ 1 +  
 \ggcf \left( 12.8587 -3.4333 \csw -1.4288 \csw^2  
 \right)  + O(g^4) \right].  
 \end{equation}  
 
 In perturbation theory the quark propagator has a pole in the
 complex momentum plane at $p^2 = - m_{\rm pole}^2$. We look for a 
 value of $p^2$ where $S^{-1}(p,m)$ in eq.~(\ref{sig01def}) has a 
 zero eigenvalue. Using eqs.~(\ref{r0ans}) and (\ref{r1ans}) 
 gives for the pole mass
 \begin{eqnarray} 
  \lefteqn {
 m_{\rm pole} \left( 1  + \ggcf [1 + a m ( \csw - 1) ]
                6 \ln (a m_{\rm pole}) \right) }  \\
 & = & \; m \; \left( 1 + \ggcf \left( 16.95241 + 7.73792 \csw 
  - 1.38038 \csw^2 \right) \right)   \nonumber \\ 
 \times \Bigg( 1  & - & \frac{a m }{2}\left[ 1 + \ggcf \left( 
 12.32005  + 18.70718 \csw 
  - 8.23318 \csw^2 \right)\right] \Bigg) \nonumber .
 \label{polemass} 
 \end{eqnarray} 
 Note that $ m_{\rm pole}$ is gauge invariant, as it should
 be~\cite{kronf}. At $\csw = 1$ the pole mass is given by
 \begin{eqnarray} 
  && m_{\rm pole} \Big( 1  +  \ggcf  
                6 \ln (a m_{\rm pole}) \Big) \\
  && \qquad \qquad = m \,  \left( 1 + \ggcf 23.30995 \right)  
  \left(1 - \frac{a m }{2}\left( 1 + \ggcf  
  22.79406 \right) \right) \nonumber \\
  && \qquad \qquad = m_\star \,  \left( 1 + \ggcf 23.30995 \right)
 \nonumber .
 \end{eqnarray} 
 The pole mass becomes a function of $m_\star$, with the same 
 value of $b_m$ as in eq.~(\ref{propcoef}). 
 The unwanted $ a m \ln (a m_{\rm pole}) $ term vanishes, and
 so the logarithm  has the same coefficient as in the continuum. 
 We will see that this is always the case, that when $\csw \ne 1$ the
 coefficients of the logarithm are changed by an amount proportional to
 $a m$, but at $\csw = 1$ all logarithmic terms have their 
 correct values. 
 
 We define our renormalisation constants $Z$ in two different 
 renormalisation schemes, $\MSB$, and a momentum 
 subtraction scheme, which we will call ${\rm MOM}$. In both cases we  
 define the $Z$s in terms of the improved fermion propagator $S_\star$.  
  
 The $\MSB$ renormalisation constants are defined from  
 \begin{equation} 
 S_\star(p, m_\star)  
 = Z_2^{\msb}(\mu^2 ) 
       S_{\msb}\left(p,  m^{\msb} \right) 
 = Z_2^{\msb}(\mu^2 ) 
       S_{\msb}\left(p,  Z_m^{\msb}(\mu^2 ) m_\star \right) ,
 \label{szszs}
 \end{equation}  
 where  $ S_{\msb}$ is the continuum fermion propagator 
 calculated perturbatively in the $\MSB$ scheme at the scale $\mu$:
\begin{eqnarray}
 \lefteqn{ 
 S^{-1}_{\msb}(p,m_\msb) = {\rm i} \, \pslash + m_\msb} \\
  &-& {\rm i}\,\pslash \, \ggcf \left[ \,
       - \alpha
   + \alpha \, \ln \left( \frac{p^2 + m_\msb^2}{\mu^2} \right)
    +  \alpha \frac{m_\msb^2}{p^2}
 \left( 1 - T\left(p^2/m_\msb^2 \right) \right)
      \right]  \nonumber \\
      & - &  m_\msb  \ggcf \left[ 
     -4 - 2\,\alpha
   + ( 3 + \alpha  ) \ln \left( \frac{p^2 + m_\msb^2}{\mu^2} \right) 
  + ( 3 + \alpha )  T\left(p^2/m_\msb^2 \right)
  \right] . \nonumber 
 \end{eqnarray}
  
 In the $\MSB$ scheme at the scale $\mu$  
 we find for the renormalisation coefficients $Z^{\msb}_2$  
 and $Z^{\msb}_m$ as defined in eq.~(\ref{szszs}):   
 \begin{eqnarray}  
 Z_2^{\msb}( \mu^2 ) &=&   
  1+ \ggcf  
 \Big[ 2 \, \alpha \, \ln(a \mu) + 16.64441  - 2.24887\csw \nonumber \\
 && \qquad \qquad - 1.39727 \,\csw^2  
 - 3.79201 \, \alpha \, \Big] \,, \\  
 Z_m^{\msb} ( \mu^2 ) 
   &=& 1-\frac{g^2}{16\pi^2}\;C_F\;\Big[6 \ln (a\mu)   
  - 12.95241 \nonumber \\
 && \qquad \qquad - 7.73792\csw + 1.38038\csw^2\Big]\, .  
 \label{ZMS}  
 \end{eqnarray}  

 In ${\rm MOM}$ we define the $Z$s at the subtraction scale $M$ through 
 \begin{equation} 
 S_\star(p, m_\star) =  \frac{Z_2^{\rm MOM}(M^2 )}{  
       {\rm i} \pslash +  Z_m^{\rm MOM}(M^2 ) m_\star }   
 \end{equation}  
 when $p^2 = M^2$. This implies that 
 \begin{eqnarray} 
 -{\rm i} \frac{1}{4 N_c} {\rm Tr} 
 \left[ \pslash S_\star^{-1}(p, m_\star) \right] 
 &=& \frac{p^2}{Z_2^{\rm MOM}(M^2 )} \,, \\
 \frac{1}{4 N_c} {\rm Tr}  \left[ S_\star^{-1}(p, m_\star) \right] 
  &=& m_\star \frac {Z_m^{\rm MOM}(M^2 )}{Z_2^{\rm MOM}(M^2 )} \;. 
 \end{eqnarray} 
 The advantage of the ${\rm MOM}$ scheme is that all the   
 quantities involved can be calculated on the lattice,
 so it can be used non-perturbatively too.  
 This is different from the $\MSB$ scheme, where we need to compare
 with a continuum quantity which we can only find perturbatively.  
  
 The  $Z$s in the ${\rm MOM}$ scheme are not simple as in 
 the $\MSB$ scheme, they still have mass and 
 gauge dependences which cancel in the  $\MSB$ case:  
 \begin{eqnarray}  
 Z_2^{\rm MOM}( M^2 ) &=&   
  1+ \ggcf   
  \Big[ 16.64441 - \alpha     
      -  2.24887\csw - 1.39727\csw^2  
  \\ 
 & & \qquad  +\,\alpha \,L(a M,a m_\star)  
  + \alpha \starminT 
      \Big] + O(a) \nonumber , \\  
 Z_m^{\rm MOM}( M^2 ) &=&   1+\ggcf \;\Big[ 
 5.57638 + \alpha   
 + 7.73792\csw - 1.38038\csw^2  
 \\  
 && \qquad  -3  \, L( a M , a m_\star )  
 - (3 + \alpha)  \starTterm \nonumber \\
 && \qquad \qquad + \alpha \starminT   
 \Big]+ O(a) \nonumber .
\label{Zmom} 
 \end{eqnarray}  
  Note however that $Z_m^{\rm MOM}$ becomes gauge independent 
 when the fermion is on-shell, i.e. at the point $M^2 = -m_{\rm{pole}}^2$. 

 The dependence on the lattice spacing $a$ and  
 clover coefficient $\csw$ is the same in the  $\MSB$ and  
 ${\rm MOM}$ schemes, so that the ratio $Z^\msb/Z^{\rm MOM}$ is independent  
 of $\csw$. This is as it should be, because the ratio 
 $Z^{\msb}/Z^{\rm MOM}$ 
 is simply the conversion factor between the two schemes, which can be 
 calculated in the continuum, and so should not refer to the lattice 
 in any way.  
   
 \section{Renormalisation and improvement of quark bilinear operators }  
   
\subsection{Method \label{opmethod}} 
 
  We are interested in calculating the $Z$ factors and improvement
 coefficients for 
 quark bilinear operators. Let us first set out our notation for  
 a general operator. We consider forward matrix elements 
 only, so improvement operators proportional to  
 a total derivative will be dropped.   
 
   All the operators in Sect.~\ref{basis}  have the form 
 \begin{equation}
 \cal{O}^{\rm imp} =
 \psib O \psi + a m c_0 \psib O \psi +  
 a \sum_{k=1}^n  c_k \psib Q^k \psi ,
 \label{Gamimp0}
 \end{equation}
 where $  \psib O \psi$ is the original unimproved operator, and
 the $ \psib Q^k  \psi$ are 
 operators with the same symmetries as the original,
 but dimension one higher. Explicit expressions for the $Q^k$ can 
 be found in Table~\ref{basistab}. For example, $Q^1$ for the
 scalar operator is $-\Dlrsl/2$. 

 We define the flavour non-singlet Green's function in the 
 usual way:
 \begin{eqnarray} 
 \lefteqn{G^{ \calO^{\rm imp} } (p,m;c_0,c_1,\cdots,c_n)
 = \frac{1}{V} \sum_{ x, x',\, y, y' } e^{-{\rm i} p \cdot (x - x') }}
 \nonumber \\
 &&\times 
 \left\langle \psi(x) \psib(y) \Big(O + a m c_0 O + a c_1 Q^1 
 + \cdots \Big)_{y,y'} 
 \psi(y')  \psib(x') \right\rangle_{ \psib, \psi , A} 
 \nonumber \\ &=& 
 \frac{1}{V} \sum_{ x, x' } e^{-{\rm i} p \cdot (x - x') } 
 \nonumber \\
 && \times  
 \left\langle \Big[ M^{-1} (O + a m c_0 O + a c_1 Q^1 + \cdots + a c_n Q^n ) 
 M^{-1} \Big]_{x,x'}  \right\rangle_A  \nonumber  
 \\ &\equiv& \FT
 \left\langle  M^{-1} (O + a m c_0 O + a c_1 Q^1 + \cdots + a c_n Q^n ) 
 M^{-1}  \right\rangle_A ,   
 \label{psi4}
 \end{eqnarray} 
  where $M$ is the fermion matrix, and $\FT$ denotes the Fourier transform
 from position to momentum space. 
 The $c_0$ dependence of $ G^{ \calO^{\rm imp} }$ is simple. We can write 
 \begin{eqnarray}
 G^{ \calO^{\rm imp} } (p,m;c_0,c_1,\cdots,c_n) &=& 
 G^{ \calO^{\rm imp} } (p,m;0,c_1,\cdots,c_n) \nonumber \\
 && + a m c_0 G^{ \calO^{\rm imp} } (p,m;0,0,\cdots,0) ,
 \label{restore_c0} 
 \end{eqnarray} 
 so we only need to give expressions for $G$ for the case $c_0 = 0$.

\begin{figure}[tb]
\vspace*{-6.0cm}
\begin{center}
\epsfig{file = 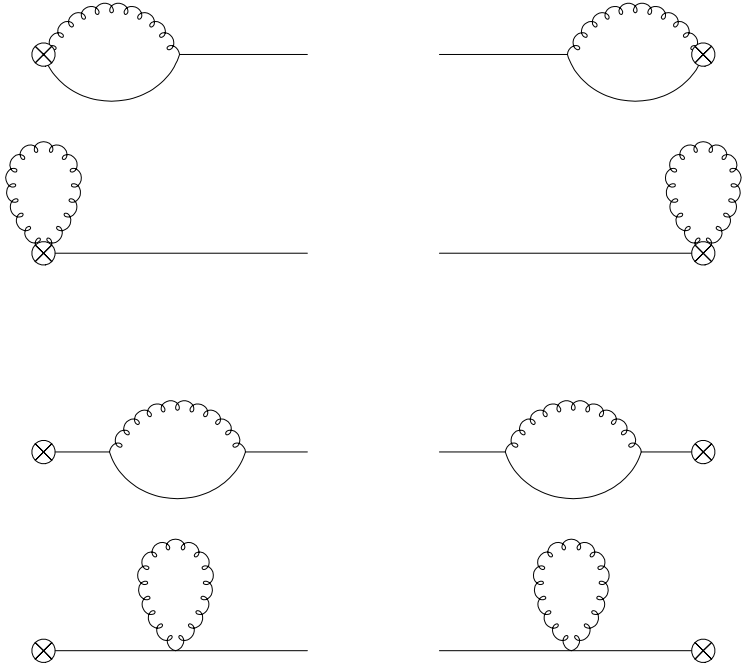, width = 13cm}
\end{center} 
 \vspace*{-2.8cm}
\caption{ The one-loop lattice Feynman diagrams needed for the contact 
 Green's function $C_\calO(p,m)$, defined in eq.~(\ref{contactgreen}).}
\label{fig:confey}
\vspace{0.5 cm}
\end{figure}

 Just as we have contact terms in the fermion propagator, we should expect 
 to see $O(a)$ contact terms arising in eq.~(\ref{psi4}) when $x = y$
 or $x' = y'$. These will give rise to a ``contact Green's function", 
 $C^\calO$, which will have to be subtracted from the operator 
 Green's function, just as we subtracted a $\delta$ function from the
 fermion propagator. This contact term is thus given by
 \begin{eqnarray}
 C^\calO(p, m) &= & 
 \frac{1}{V} \sum_{ x, x' } e^{-{\rm i} p \cdot (x - x') }
 \left\langle \Big[ M^{-1} O + O M^{-1} \Big]_{x,x'}
 \right\rangle_A \nonumber \\
 &\equiv& \FT  \left\langle M^{-1} O + O M^{-1}  \right\rangle_A. 
 \label{contactgreen} 
 \end{eqnarray}  
  Since the coefficient of $C^\calO$ will be $O(a)$, we only need
 to calculate it for the unimproved operator $\psib O \psi$, 
 and we only need the leading order in $a$. The Feynman diagrams
 needed to calculate $C^\calO$ to $O(g^2)$ are shown in  
 Fig.~\ref{fig:confey}. There is no extra calculation involved. All
 the graphs needed already occur in the perturbative expansion of the 
 operator and propagator.\footnote{A complete listing
 of the graphs can be found 
 for example in~\cite{capi1}, \cite{capi2} or~\cite{qcdsf5}.}

 Finding the contact Green's function is simple when we consider
 point operators of the type  $\psib \Gamma_i \psi$ where
 $\Gamma_i$ is any $4 \times 4$ matrix. Because there are no 
 covariant derivatives in the operator, it is unaffected
 by the averaging over gauge fields, and eq.~(\ref{contactgreen})
 simplifies to give
 \begin{equation} 
 C^{\Gamma_i}(p, m) = \Gamma_i S(p,m) + S(p,m) \Gamma_i 
 = \left\{ \Gamma_i \, ,\, S(p,m) \right\} \; .
 \label{C_Gamma} 
 \end{equation} 
 The contact Green's function are more complicated in the general case.
 Their expressions for the one-link operators are given in the Appendix
 (Sects.~(\ref{sectunpold}) to (\ref{sectpolu})). 

 So, by subtracting a contact term proportional to $ C^\calO (p,m)$ in a
 Green's function and by choosing appropriately the improvement 
 coefficients, an improved Green's function can be obtained.
 The resulting expression for a renormalised improved off-shell 
 Green's function is:
 \begin{eqnarray}
   G_R^\calO(p, m_R; \mu^2)  &=&  
  \frac{ Z_\calO ( \mu^2; c_1, \cdots ,c_n ) } 
 {Z_2(\mu^2) \, (1 + a m b_\psi ) } \\  
 &&\times \Big[ G^{\cal O^{\rm imp}}(p,m;c_0,c_1, \cdots , c_n ) 
  - \frac{a}{2} \lambda_\calO C^\calO(p,m)  
  \Big] . \nonumber 
 \end{eqnarray}
  The factor $1/[Z_2 (\mu^2) (1 + a m b_\psi )]$ accounts for the 
 wave-function renormalisation. 
 The Green's function in eq.~(\ref{psi4}) depends linearly on
 the $c_k$ coefficients, while the renormalised Green's function is 
 independent of the $c_k$s. From this we can deduce that
 $1/ Z_\calO ( \mu^2; c_1, \cdots , c_n )$
 must depend linearly on $c_k$ too, so we can write
 \begin{equation}
 Z_\calO ( \mu^2; c_1, \cdots , c_n) = 
 \frac { Z_\calO ( \mu^2; 0, \cdots , 0 ) }
 { 1 + \sum_{k=1}^n \zeta_k c_k } , 
 \label{defzeta}
 \end{equation} 
 where the $\zeta$s are coefficients depending on $g^2$. 
 At the one-loop level, using the fact that $Z = 1 + O(g^2)$ and that
 the $\zeta$s are  $O(g^2)$, we can write 
 \begin{equation}
  Z_\calO ( \mu^2; c_1, \cdots , c_n) =  Z_\calO ( \mu^2; 0, \cdots , 0 )
 -  \sum_{k=1}^n \zeta_k c_k + O(g^4)  \, .
 \label{Zc_oneloop} 
 \end{equation} 
 Our final formula for the renormalised  and improved Green's
 function is
 \begin{eqnarray}
   G_R^\calO(p, m_R; \mu^2)  & =&   
  \frac{1}{Z_2(\mu^2) \, (1 + a m b_\psi ) }  
   \frac{Z_\calO (\mu^2; 0, \cdots ,0 )}{1 + \sum_k \zeta_k c_k}  
 \nonumber \\
 && \times \left[ G^{\calO^{\rm imp} }(p,m;c_0, c_1, \cdots, c_n)
   - \frac{a}{2} \lambda_\calO C^\calO(p,m) 
 \right] .
 \label{renimp_3pt}
 \end{eqnarray}

   The improvement coefficients $b, c, \lambda$ and $\zeta$ are independent
 of the renormalisation scheme, while the renormalisation constants
 $Z$ are in general scheme dependent. Therefore it can be useful to
 split the renormalisation and improvement into two stages, and
 to define ``improved bare" Green's functions by
 \begin{eqnarray}
   G_\star^\calO(p, m_\star )   &=&    
  \frac{1}{ 1 + a m b_\psi  } \;  
  \frac{1}{1 + \sum_k \zeta_k c_k}  \label{imp0_3pt} \\
  && \times \left[ G^{\calO^{\rm imp} }(p,m;c_0, c_1, \cdots, c_n) 
      - \frac{a}{2} \lambda_\calO C^\calO(p,m) \right] \,, \nonumber 
 \end{eqnarray}
 where $ m_\star =  m (1 + a m b_m  )$, using the same value 
 for $b_m$ as found from the fermion propagator 
 (eq.~(\ref{propcoef})). As in the propagator section, 
 we will use the suffix $\star$ to denote bare quantities which have 
 been improved to $O(a)$.

 The second step is then to renormalise
 this improved Green's function multiplicatively,
 \begin{equation}
 G_R^\calO(p, m_R ; \mu^2) = 
 \frac{Z_\calO (\mu^2;0,\cdots,0)}{Z_2(\mu^2) }
 G_\star^\calO(p, m_\star). 
 \label{GZdef} 
 \end{equation} 

\begin{figure}[tbh]
\vspace*{-5.0cm}
\begin{center}
\epsfig{file = 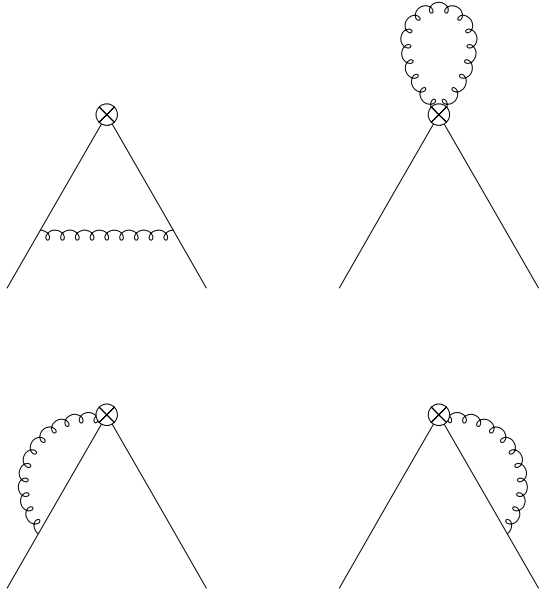, width = 12cm}
\end{center} 
 \vspace*{-3.2cm}
\caption{ The one-loop lattice Feynman diagrams needed for the amputated 
 Green's function (vertex function), 
 $\Lambda^{\calO^{\rm imp}}(p,m; c_0, c_1, \cdots ,c_n)$,
 as defined in eq.~(\ref{ampu}). }
\label{fig:lambfey}
\vspace{0.5 cm}
\end{figure}

 It is useful to write corresponding equations for amputated Green's
 functions too. We define the amputated Green's function
 $\Lambda$ in the standard way:
 \begin{equation}
  \Lambda^{\calO^{\rm imp}} (p, m; c_0, c_1, \cdots ,c_n) \equiv 
 S^{-1}(p,m) G^{\calO^{\rm imp}} (p, m; c_0, c_1, \cdots ,c_n) S^{-1}(p,m) ,
 \label{ampu} 
 \end{equation} 
 where $S(p,m)$ is the full fermion propagator. 
 The amputation of eq.~(\ref{ampu}) removes all fermion self-energy
 diagrams from the perturbative expansion of $\Lambda$. 
 The one-loop Feynman diagrams for $\Lambda$ are shown in
 Fig.~\ref{fig:lambfey}.

   The improved amputated Green's function, 
 $\Lambda_\star$, is naturally defined by  
 \begin{equation}
 \Lambda^\calO_\star (p, m_\star) \equiv  S_\star^{-1}(p,m_\star) 
 G^\calO_\star (p, m_\star) S_\star^{-1}(p,m_\star) . 
 \end{equation} 
   From eq.~(\ref{GZdef}) we obtain the renormalised amputated
 Green's function:  
 \begin{equation}
 \Lambda_R^\calO(p, m_R ; \mu^2) = 
  Z_2(\mu^2) Z_\calO (\mu^2;0,\cdots,0)  
 \Lambda_\star^\calO(p, m_\star). 
 \label{GLdef} 
 \end{equation} 
 
 Substituting eq.~(\ref{imp_inv})
 and eq.~(\ref{imp0_3pt}) and using eq.~(\ref{ampu}) we find 
 \begin{eqnarray}
 \Lambda_\star^\calO(p, m_\star) &=& 
 \frac{1 + a m b_\psi}{1 + \sum_k \zeta_k c_k} \Big[
 \Lambda^{\calO^{\rm imp}}(p, m ; c_0, c_1, \cdots, c_n ) \nonumber \\
 && -\frac{a}{2} \lambda_\calO
  S^{-1}(p, m)  C^\calO(p,m)  S^{-1}(p, m) 
  \nonumber \\ &&
 +\frac{a}{2} \lambda_\psi
 \left\{ S^{-1}(p, m) , 
 \Lambda^{\calO^{\rm imp}} (p, m; c_0, c_1, \cdots, c_n) \right\} 
 \Big] \; . 
 \label{imp0_Lambda}
 \end{eqnarray}
  Similarly to what was done in the case of the propagator, one can now 
  expand $\Lambda^\calO$, $S^{-1}$ and $C^\calO$ in powers of $a$ 
  and thus obtain a non-linear condition analogous to eq.~(\ref{sig1eqn}), 
  from which the improvement coefficients can be derived.  
 In the case of the local operators $\psib \Gamma_i \psi$, the 
 expression eq.~(\ref{C_Gamma}) for the contact term means that we
 can write the expression for $\Lambda_\star$ in the simpler form 
 \begin{eqnarray}
 \lefteqn{ 
 \Lambda_\star^{\Gamma_i}(p, m_\star)  =  
 \frac{1 + a m b_\psi}{1 + \zeta_1^{\Gamma_i} c_1^{\Gamma_i}} \Big[
 \Lambda^{\Gamma_i^{\rm imp}}(p, m ; c_0, c_1 ) } \nonumber \\
 && -\frac{a}{2} \lambda_{\Gamma_i}
  \left\{ S^{-1}(p, m) , \Gamma_i \right\}  
 +\frac{a}{2} \lambda_\psi
 \left\{ S^{-1}(p, m) , 
 \Lambda^{\Gamma_i^{\rm imp}} (p, m; c_0, c_1 ) \right\} 
 \Big] \, . 
 \label{imp0_Lambda_Gam}
 \end{eqnarray}

  From eq.~(\ref{imp0_Lambda}) we can see that the two improvement  
  coefficients $\lambda_\psi$ and $\lambda_\calO$ associated with
  the contact terms are only needed  for off-shell improvement, because
  the inverse propagator $S^{-1}(p,m)$  vanishes on-shell.  
 
 \subsection{Results for point quark operators \label{point_sect}}  

 We shall now calculate the matrix elements of all point operators up 
 to $O(g^2 a)$ including the finite terms. This goes beyond
 the work of Heatlie et al.~\cite{heatli}, who only considered the 
 $g^2 a\ln a$ terms. Including all $O(g^2 a)$ terms will enable us 
 to compute the improvement coefficients to $O(g^2)$.

 The calculations are carried out for arbitrary $m^2/p^2$, not just for 
 $m^2 \ll p^2$ as in our previous papers~\cite{caplat97,qcdsf5,stlouis}.
  We give the results for the amputated Green's functions.

 In this section we show how the improvement coefficients and 
 renormalisation constants are calculated in one particular case, 
 that of the scalar operator
 \begin{equation}
 \left( \calO^S \right)^{\rm imp} = 
 \left(\psib\psi\right)^{\rm imp} = 
 (1 + a\,m\,c_0)  \psib\psi -   
 \frac{1}{2} a c_1 \psib\Dlrsl\psi .  
 \end{equation}
 The Green's functions and results for the other  
 operators are given in the Appendix. We consider forward matrix elements,
 therefore we drop the total derivative terms in the improved bases
 (although in the scalar case this does not make any difference), 
 which now all have the form
\begin{equation}
 \calO_\Gamma^{\rm imp} = 
 \psib \Gamma \psi -  \frac{1}{4} a
 c_1^\Gamma \psib \{ \Dlrsl , \Gamma \} \psi
 + a m c_0^\Gamma \psib \Gamma \psi . 
\label{Gamimp2}
\end{equation}

 The one-loop expression for the amputated scalar Green's function  
 up to $O(a)$ is:
\begin{eqnarray}
 \lefteqn{ \Lambda^S(p,m; 0, \cscl) = 
     1 - a\,{\rm i}\pslash\,\cscl  }
 \nonumber \\ 
 &+&\ggcf \, \Big[-11.06803 + 2\,\alpha + 9.98679 \csw + 0.01689 \,{\csw^2} 
 \nonumber \\ &&
   + \,\cscl ( - 19.17181 + 13.80068 \csw - 3.53833\,{\csw^2} ) 
 \nonumber \\ &&
      - ( 3 + \alpha ) \,L(a p, a m) - 3 \, ( 3 + \alpha ) \,\Tterm \Big]
 \nonumber \\ 
 &+& {\rm i} \pslash\, \ggcf \,  4\,\alpha\, \frac{m}{p^2} \,
      \Big[  - 1 + \Tterm  \Big]
 \nonumber \\ 
 &+& a\,{\rm i}\pslash\,\ggcf \Big[ 6.34664 - 0.14375\,\alpha 
      + 1.48503\csw - 1.28605\,{\csw^2}
 \nonumber \\ &&
   +\cscl (  11.65102  
  - \alpha - 1.41422\csw + 0.78465\,{\csw^2} )  
 \nonumber \\ &&
      + ( {\frac{3}{2}} + \alpha + \alpha\,\cscl +
         {\frac{3}{2} \csw} ) \,L(a p, a m) 
 \nonumber \\ &&
   + 3\, \left( \alpha + 3\csw \right) \Tterm
     - 4\,\alpha\,\frac{m^2}{m^2 + p^2}
 \nonumber \\ &&
  + \left( \frac{15}{2} + 10\,\alpha - 3\,\alpha\,\cscl
      - \frac{15}{2}\csw \right) \,\frac{m^2}{p^2}
    \left( 1 -  \Tterm \right)
 \Big]
 \\ 
 &+& a\,m\,\ggcf \Big[ 31.06826 - 4.14375\,\alpha
      - 33.97148\csw + 3.04688\,{\csw^2}  
 \nonumber \\ &&
 + \cscl ( - 6.08204  + 2.20074\csw  + 1.44647\,{\csw^2} )
 \nonumber \\ &&
 + ( 9 + \alpha - 6\csw ) \,L(a  p, a  m)  
    - 2 \, ( 3  + \alpha ) \,\frac{m^2}{m^2 + p^2}
 \nonumber \\ &&
       + 2 \, ( 12 + 5\,\alpha - 3\,\cscl -
              \alpha\,\cscl - 3\csw ) \, \Tterm   
     \Big] \nonumber .
\end{eqnarray}
 We only need to give the expression for $\Lambda^S$ when $c_0 = 0$, 
 since the full expression with non-zero $c_0$ can be recovered by using 
 eq.~(\ref{restore_c0}). 

 To improve the Green's functions, we need to 
  choose the improvement coefficients so that 
 all $O(a)$ terms in the expressions  eq.~(\ref{imp0_3pt}) or 
 eq.~(\ref{imp0_Lambda}) vanish. It is not immediately obvious that
 this will be possible, because there are many more $O(a)$ terms than
 there are improvement coefficients, and therefore more equations
 to be satisfied than there are unknowns. For general $\csw$ we can
 not satisfy all the equations, we can only remove all $O(a)$ effects if 
 $\csw = 1 +O(g^2)$. In this case we can derive perturbative expressions  
 for the improvement coefficients. The results are 
\begin{eqnarray} 
 c_0^S &=&  1 - \cscl + \ggcf ( 22.79406 - 8.45146 \cscl) ,\nonumber \\ 
 \lambda_S &=& 1 - \cscl + \ggcf (16.39210 - 10.88629 \cscl) ,\nonumber \\    
 \zeta_1^S &=& - \ggcf  8.90946 .  
\label{impco_scalar} 
\end{eqnarray}  
 All three improvement coefficients are gauge invariant. 
 There is one free parameter in this system of equations. The improvement 
 coefficient $\cscl$ can take any value, but once it is chosen, the values 
 of the other improvement coefficients are fixed. This freedom comes from 
 an equation of motion, which allows us to compensate for a change
 in one of the improvement coefficients by adjusting the other
 two coefficients.         
 For example, there is a particularly interesting value of $\cscl$ 
 where $\lambda_S$   
 vanishes, which means that the scalar three-point function contains  
 no contact terms, and so even off-shell it is simply renormalised 
 by a multiplicative factor. This value of $\cscl$ is  
\begin{equation}  
 {\tilde c}_1^S = 1 + \ggcf 5.50582 + O(g^4) .  
 \label{c1tild_scalar} 
\end{equation}  
 The improvement coefficients $c_0^S$ and $\lambda_S$ are only defined 
 at $\csw = 1$, but $\zeta_1^S$ can also be defined for general $\csw$ values.  
 The result is  
\begin{equation}    
 \zeta_1^S = \ggcf (-19.17181 + 13.80068 \csw - 3.53833 \csw^2 ) .
\end{equation}  
 All these results are gauge invariant.

\renewcommand{\arraystretch}{2}

\begin{table}[Htb]
\vspace{0.5 cm}
\begin{center}
\begin{tabular}{|c||l|}
\hline
 &  \multicolumn{1}{c|}{$Z^\msb(\mu^2; 0)$}\\ \hline 
 $S$ & $ 1 + \ggcf ( -12.95241 - 7.73792 \csw +  1.38038 \csw^2  
 + 6 \ln  a \mu )$ \\
 $P$  & $ 1 + \ggcf ( -22.59544 +2.24887 \csw -  2.03602 \csw^2 
 + 6 \ln  a \mu )$ \\
 $V$  & $ 1 + \ggcf ( -20.61780 + 4.74556 \csw + 0.54317 \csw^2 )$ \\
 $A$  & $ 1 + \ggcf ( -15.79628 - 0.24783 \csw + 2.25137 \csw^2)$\\ 
 $H$  & $ 1 + \ggcf ( -17.01808 + 3.91333 \csw + 1.97230 \csw^2 
 -2 \ln  a \mu )$ \\
 \hline 
\end{tabular} 
\end{center}
\vspace{0.5 cm}
\caption{The renormalisation constants $Z^\msb$ for the
 point operators (with no improvement term added, i.e. 
 $c_1^\Gamma = 0$).}\label{zmstab}
\vspace{0.9 cm}
\end{table} 
 
\begin{table}[Htb]
\begin{center}
\begin{tabular}{|c||l|}
\hline
 &  \multicolumn{1}{c|}{$\zeta_1^\Gamma$}\\ \hline 
 $S$ & $ \ggcf ( -19.17181 + 13.80068 \csw - 3.53833 \csw^2 )$ \\ 
 $V$ & $ \ggcf ( -9.78635 + 3.41640 \csw +  0.88458 \csw^2 )$ \\
 $A$ & $ \ggcf ( -19.37225 + 10.31673 \csw -  0.88458 \csw^2 )$ \\ 
 $H$ & $ \ggcf ( -16.24376 + 6.85531 \csw +  0.58972 \csw^2 )$ \\
 \hline 
\end{tabular} 
\end{center}
\vspace{0.5 cm}
\caption{The improvement coefficients $\zeta_1^\Gamma$. These give 
 the renormalisation constants when $c_1^\Gamma$ is non-zero, see
 eq.~(\ref{Zc_oneloop}).  }\label{zeta}
\vspace{1.0 cm}
\end{table}

\begin{table}[Htb]
\vspace{0.5 cm}
\begin{center}
\begin{tabular}{|c||l|}
\hline
  & \multicolumn{1}{c|}{$ c_0^\Gamma $}\\
\hline 
$S $&$ 1 - \cscl + \ggcf ( 22.79406 - 8.45146 \, \cscl )$ 
 \\ 
$V $&$ 1 - \cmu + \ggcf ( 18.14912 - 6.96476 \, \cmu )$
 \\ 
$A $&$ 1 - \cax + \ggcf ( 18.02539 - 13.36028 \, \cax )$
 \\ 
$H $&$ 1 - \ct + \ggcf ( 16.47708 - 12.86471 \, \ct )$ 
 \\  \hline 
\end{tabular} 
\end{center}
\vspace{0.5 cm}
\caption{The improvement coefficients $c_0^\Gamma$ 
 for general $c_1^\Gamma$ at $\csw =1$.}\label{bc1}
\vspace{1.0 cm}
\end{table} 

\begin{table}[Htb]
\begin{center}
\begin{tabular}{|c||l|}
\hline
 &\multicolumn{1}{c|}{ $\lambda_\Gamma$ }\\ 
\hline 
$S $&$ 1 - \cscl + \ggcf (16.39210 - 10.88629  \, \cscl )$ 
 \\ 
$V $&$ 1 - \cmu + \ggcf (13.71395 - 2.29988 \, \cmu )$
 \\ 
$A $&$ 1 - \cax + \ggcf ( 11.46014 - 8.37649 \, \cax )$
 \\ 
$H $&$ 1 - \ct + \ggcf ( 10.56742 - 5.51435 \, \ct )$ 
 \\  \hline 
\end{tabular} 
\end{center}
\vspace{0.5 cm}
\caption{The improvement coefficients $\lambda_\Gamma$ 
 for general $c_1^\Gamma$ at $\csw =1$.}\label{lamc1}
\vspace{1.0 cm}
\end{table} 

 \begin{table}[Htb]
 \begin{center}
 \begin{tabular}{|c||r|l|}
 \hline 
 & \multicolumn{1}{c|}{${\tilde c}_0^\Gamma $}
 & \multicolumn{1}{c|}{$ {\tilde c}_1^\Gamma $} \\ 
 \hline
 $S$ &$\ggcf \, 8.83678$ & $1 + \ggcf \,  5.50582 $\\
 $P$ &$-\ggcf \, 5.21443$ & \multicolumn{1}{c|}{--}\\
 $V$ &$-\ggcf \, 0.22971$ & $1 + \ggcf \, 11.41407 $\\
 $A$ &$\ggcf \, 1.58146$ & $1 + \ggcf \, 3.08365 $\\
 $H$ &$-\ggcf \,1.44071 $& $1 + \ggcf \, 5.05307 $\\
 \hline  
 \end{tabular} 
 \end{center}
 \vspace{0.5 cm}
 \caption{The improvement coefficients $c_0^\Gamma$  
  and $c_1^\Gamma$ when the contact term
  $\lambda_\Gamma$ is chosen to be zero.}\label{c1tild}
\vspace{1.0 cm}
 \end{table} 
  
  We calculate the continuum Green's functions 
 (needed for the $Z^\msb$ factors) in the
 $\MSB$ (minimal subtraction) scheme.  In this paper we 
  use a totally anticommuting $\gamma_5$, even when $d \neq 4$.  
 For the scalar Green's function the result in the $\MSB$ scheme is
 \begin{eqnarray}
 \Lambda^S_\msb(p,m_\msb) &=& 1 
  +  \ggcf \, \Big[ 4 + 2\,\alpha 
  - ( 3 + \alpha ) \, \ln \left( \frac{p^2 + m_\msb^2}{\mu^2} \right)
 \nonumber \\ && \qquad \qquad
   - 3 \, ( 3 + \alpha ) \, T\left( p^2 / m_\msb^2 \right) \Big]
 \nonumber \\ 
 &+& {\rm i} \frac { m_\msb \, \pslash }{p^2} \, \ggcf \,  4\,\alpha\, 
      \left[ - 1 +  T\left( p^2 / m_\msb^2 \right)  \right] .
\end{eqnarray}
 We can now calculate $Z^\msb$ which is defined by 
 \begin{equation} 
 Z_2^\msb(\mu^2) Z_S^\msb(\mu^2; c_1^S) \Lambda_S(p,m; 0, c_1^S) 
 = \Lambda^\msb_S(p,m_\msb) \, . 
 \label{def_zmsb} 
 \end{equation}  
 The result is 
 \begin{eqnarray} 
  Z_S^\msb(\mu^2; c_1^S)
 &=& \frac{ 1 + \ggcf ( -12.95241 - 7.73792 \csw +  1.38038 \csw^2
 + 6 \ln  a \mu )} 
 {1 +  \ggcf ( -19.17181 + 13.80068 \csw - 3.53833 \csw^2 ) c_1^S } 
 \nonumber \\ &=& 
  \frac{ Z_S^\msb(\mu^2; 0)}{1 + \zeta_1^S c_1^S } \;. 
 \end{eqnarray} 

  In the MOM scheme we define the $Z$ for an operator ${\cal O}$ by 
  \begin{equation} 
   Z^{\rm MOM}_2(M^2) Z^{\rm MOM}_{\cal O}(M^2)  
  {\rm Tr}\left[ (\Lambda^{\rm Born}_{\cal O}(p) )^\dagger
  \Lambda_{\cal O}(p)\right] 
  =  {\rm Tr}\left[ (\Lambda^{\rm Born}_{\cal O}(p) )^\dagger
  \Lambda^{\rm Born}_{\cal O}(p)\right]  
  \end{equation} 
 when $p^2 = M^2$, where $\Lambda^{\rm Born}_{\cal O}$ is the operator's
 Born term. Applying this definition to the scalar operator gives
  \begin{equation}
   Z^{\rm MOM}_2(M^2)  Z^{\rm MOM}_S(M^2; 0) = 
  \frac {4 N_c} { {\rm Tr}\left[ \Lambda^S(p,m;0,0) \right] }
  \end{equation} 
  at $p^2 = M^2$, so 
  \begin{eqnarray}
  Z^{\rm MOM}_S(M^2; 0)  = 1 &+& \ggcf \Big[ 
  -5.57638  - 7.73792 \csw  + 1.38038 \csw^2
  \nonumber \\ && 
  -\alpha + 3 \starLapam  + 3 \, (3 + \alpha) \starTterm 
  \nonumber \\ &&
  - \alpha \starminT \Big] +O(a).
  \end{eqnarray}
 
 The same procedure can be repeated for all the local operators. 
 In Tables~\ref{zmstab}-\ref{c1tild} we give the improvement coefficients
 and $\MSB$ renormalisation constants for all point operators. 
 The  $\MSB$ renormalisation constants are defined by 
 equations analogous to eq.~(\ref{def_zmsb}). The lattice
 $\Lambda$s and $\MSB$-scheme  $\Lambda$s are all given in 
 the Appendix.  
  In several of the tables there is no entry for the pseudoscalar
 operator -- this is because it has no $c_1$ improvement term, so
 the associated quantities are not defined.  
 The $Z$s in the MOM scheme are given in the Appendix. 

 The $c_0^\Gamma$ values in table \ref{bc1} agree with the values 
 given in~\cite{Sintlat97} for the case $c_1^\Gamma \equiv 0$.
 
 \subsection{Results for one-link quark operators}  

 We consider now the operators in eq.~(\ref{nonlocop}).
 We study the improvement of these operators along the same lines 
 used for the point operators. The expressions for the contact Green's 
 functions~(\ref{contactgreen}) will be more complicated, and are given
 in the Appendix. 

 From eqs.~(\ref{imp2}), (\ref{imp5}) and (\ref{imp4}), we can see that 
 in forward matrix elements a basis for the improvement is given 
 in the unpolarised case ($ \calO_{\mu\nu},\tau_3^{(6)} $
 and $ \calO_{\mu\nu},\tau_1^{(3)} $) by
 \begin{eqnarray}  
 \left( \calO_{\mu\nu} \right)^{\rm imp } & = &  
 \frac{1}{2} \left(1+ a\, m\, c_0 \right)
 \psib\gamma_\mu \Dlr_\nu \psi +   
 \frac{1}{8}\,a\,{\rm i} c_1  \sum_\lambda \psib 
 \sigma_{\mu\lambda}\left[\Dlr_\nu,\Dlr_\lambda\right]\psi \nonumber \\  
 & & -\frac{1}{8} \,a \,c_2  
  \psib\left\{\Dlr_\mu,\Dlr_\nu\right\}\psi ,    
 \end{eqnarray}  
 in the polarised case with 
 $\mu \ne \nu$ ($\calO_{\mu\nu}^5,\tau_4^{(6)}$) by
 \begin{eqnarray} 
 \left( \calO_{\mu\nu}^5 \right)^{\rm imp} & = & \frac{1}{2}  
 \left(1+ a\, m\, c_0 \right)\psib\gamma_\mu\gamma_5 \Dlr_\nu \psi     
 - \frac{1}{4}\, {\rm i} a\, c_1\, \psib\sigma_{\mu \nu}  
 \Dlr_\nu^2 \psi \nonumber\\   
 & & 
 -\frac{1}{8}\,a \,{\rm i}\,c_2 \, \sum_{\lambda \ne \mu, \nu} 
 \psib\sigma_{\mu\lambda} 
  \gamma_5\left\{\Dlr_\lambda,\Dlr_\nu\right\}\psi ,
 \end{eqnarray}
 and in the polarised case when $\mu = \nu$ 
 ($\calO_{\mu\nu}^5,\tau_4^{(3)}$) by
 \begin{eqnarray}
 \left( \calO_{\mu\nu}^5 \right)^{\rm imp} & = & \frac{1}{2}  
 \left(1+ a\, m\, c_0 \right)\psib\gamma_\mu\gamma_5 \Dlr_\nu \psi 
 -\frac{1}{8}\,a \,{\rm i}\,c_1 \, \psib\sigma_{\mu\lambda} 
  \gamma_5\left\{\Dlr_\lambda,\Dlr_\nu\right\}\psi .
 \end{eqnarray}

\renewcommand{\arraystretch}{2}

\begin{table}[b]
\vspace{0.2 cm}
\begin{center}
\begin{tabular}{|c||c|}
\hline
  & $ Z^\msb(\mu^2; 0,0) $ \\
\hline 
$ \calO_{\mu\nu},\tau_3^{(6)} $&$ 1 + 
 \ggcf (-1.27959 + 3.87297\;\csw + 0.67826\;\csw^2 -\frac{16}{3}\ln a \mu) $ 
 \\ 
$ \calO_{\mu\nu},\tau_1^{(3)} $&$ 1 + 
 \ggcf (-2.56184 + 3.96980\; \csw + 1.03973\;\csw^2 -\frac{16}{3}\ln a \mu) $
 \\ 
$ \calO_{\mu\nu}^5,\tau_4^{(6)} $&$ 1+ 
 \ggcf (-0.34512 + 1.35931\;\csw + 1.89255\;\csw^2 -\frac{16}{3}\ln a \mu) $
 \\ 
$ \calO_{\mu\nu}^5,\tau_4^{(3)} $&$ 1+ 
 \ggcf (-0.16738 + 1.24953\;\csw +  1.99804\;\csw^2 -\frac{16}{3}\ln a \mu)$
 \\  \hline 
\end{tabular} 
\end{center}
\vspace{0.5 cm}
\caption{The $\MSB$-scheme renormalisation constants 
 for the one-link operators.}\label{Z1link}
\vspace{0.2 cm}
\end{table} 

\begin{table}[Htb]
\begin{center}
\begin{tabular}{|c||c|}
\hline
  & $ \zeta_1 $ \\
\hline 
$ \calO_{\mu\nu},\tau_3^{(6)} $&$ 
 \ggcf ( -4.27417 + 1.08793 \csw) $ 
 \\ 
$ \calO_{\mu\nu},\tau_1^{(3)} $&$ 
 \ggcf ( -4.27417 + 1.08793 \csw )$
 \\ 
$ \calO_{\mu\nu}^5,\tau_4^{(6)} $&$ 
  \ggcf ( -5.61603 + 4.10778 \csw -0.26315 \csw^2 ) $
 \\ 
$ \calO_{\mu\nu}^5,\tau_4^{(3)} $&$ 
 \ggcf ( -15.31376 + 8.54773\;\csw - 0.26036\;\csw^2) $
 \\  \hline
\end{tabular} 
\end{center}
\vspace{0.5 cm}
\caption{The improvement coefficients $\zeta_1$
 for the one-link operators.} \label{zet1link}
\vspace{0.5 cm}
\end{table} 

\begin{table}[Htb]
\begin{center}
\begin{tabular}{|c||c|}
\hline
  & $ \zeta_2 $ \\
\hline 
$ \calO_{\mu\nu},\tau_3^{(6)} $&$ 
 \ggcf ( -9.40584 + 4.60327 \;\csw + 0.46669 \;\csw^2) $ 
 \\ 
$ \calO_{\mu\nu},\tau_1^{(3)} $&$ 
 \ggcf ( -6.67330 + 4.53710 \;\csw + 0.44621 \;\csw^2 )$
 \\ 
$ \calO_{\mu\nu}^5,\tau_4^{(6)} $&$ 
  \ggcf ( -8.29791 + 4.21724 \;\csw -0.49384 \;\csw^2 ) $
 \\ 
$ \calO_{\mu\nu}^5,\tau_4^{(3)} $& $-$
 \\  \hline 
\end{tabular} 
\end{center}
\vspace{0.5 cm}
\caption{The improvement coefficients $\zeta_2$ for the one-link
 operators.}
\label{zet2link}
\vspace{0.5 cm}
\end{table} 

\begin{table}[Htb]
\begin{center}
\begin{tabular}{|c||c|}
\hline
  & $ c_0 $ \\
\hline 
$ \calO_{\mu\nu},\tau_3^{(6)} $&$ 1 - c_2 + 
 \ggcf ( 16.34500 - 12.24534 \, c_2 ) $ 
 \\ 
$ \calO_{\mu\nu},\tau_1^{(3)} $&$ 1 - c_2 + 
 \ggcf ( 17.20377 - 8.69045 \, c_2 )$
 \\ 
$ \calO_{\mu\nu}^5,\tau_4^{(6)} $&$  1 - c_2 +
 \ggcf ( 16.28373 -  10.73103 \, c_2 ) $
 \\ 
$ \calO_{\mu\nu}^5,\tau_4^{(3)} $&$ 1 - c_1 + 
 \ggcf ( 16.95724 - 11.33434 \, c_1 ) $ 
 \\  \hline 
\end{tabular} 
\end{center}
\vspace{0.5 cm}
\caption{The improvement coefficients $c_0$ for general $c_2$ at $\csw =1$
 for the one-link operators.}\label{fc0tab}
\vspace{0.5 cm}
\end{table} 

\begin{table}[Htb]
\begin{center}
\begin{tabular}{|c||c|}
\hline
  & $ c_1 $ \\
\hline 
$ \calO_{\mu\nu},\tau_3^{(6)} $&$ c_2 + O(g^2) $ 
 \\ 
$ \calO_{\mu\nu},\tau_1^{(3)} $&$ c_2 + O(g^2) $
 \\ 
$ \calO_{\mu\nu}^5,\tau_4^{(6)} $&$ c_2 + 
 \ggcf ( 1.18321 + 0.84682 \,c_2 ) $
 \\ 
$ \calO_{\mu\nu}^5,\tau_4^{(3)} $&$ - $ 
 \\  \hline 
\end{tabular} 
\end{center}
\vspace{0.5 cm}
\caption{The improvement coefficients $c_1$ 
 for general $c_2$ at $\csw =1$ for the one-link operators.}\label{fc1tab}
\vspace{0.5 cm}
\end{table} 

\begin{table}[Htb]
\begin{center}
\begin{tabular}{|c||c|}
\hline
  & $ \lambda_\calO $ \\
\hline 
$ \calO_{\mu\nu},\tau_3^{(6)} $&$ 1 - c_2 +
  \ggcf ( 8.13135 - 3.04784 \,c_2 ) $ 
 \\ 
$ \calO_{\mu\nu},\tau_1^{(3)} $&$ 1 - c_2 + 
  \ggcf ( 9.28735 - 0.52613 \,c_2 ) $
 \\ 
$ \calO_{\mu\nu}^5,\tau_4^{(6)} $&$ 1 - c_2 +
 \ggcf ( 8.48845  -1.84428 \,c_2) $
 \\ 
$ \calO_{\mu\nu}^5,\tau_4^{(3)} $&$ 1 - c_1 + 
 \ggcf ( 7.96628 - 3.12321 \,c_1 ) $ 
 \\  \hline 
\end{tabular} 
\end{center}
\vspace{0.5 cm}
\caption{The improvement coefficients $\lambda_\calO$ for general
 $c_2$ at $\csw =1$ for the one-link operators.}\label{flambdatab}
\vspace{0.5 cm}
\end{table} 

\begin{table}[Htb]
\begin{center}
\begin{tabular}{|c||c|c|c|}
\hline
  & $ {\tilde c}_0 $ & $ {\tilde c}_1 $ & $ {\tilde c}_2 $ \\
\hline 
$ \calO_{\mu\nu},\tau_3^{(6)} $&$ -\ggcf 0.98385  $
 &$ {\tilde c}_2 + O(g^2) $&$ 1 + \ggcf 5.08351  $ 
 \\ 
$ \calO_{\mu\nu},\tau_1^{(3)} $&$ -\ggcf 0.24789  $
 &$ {\tilde c}_2 + O(g^2) $&$ 1 + \ggcf 8.76121  $
 \\ 
$ \calO_{\mu\nu}^5,\tau_4^{(6)} $&$ -\ggcf 1.09147 $&$ 
 1 + \ggcf 8.67421  $&$ 1 + \ggcf 6.64417 $
 \\ 
$ \calO_{\mu\nu}^5,\tau_4^{(3)} $&$ 
 +\ggcf 0.77983 $&$ 1 + \ggcf 4.84307 $&$ -$
 \\  \hline 
\end{tabular} 
\end{center}
\vspace{0.5 cm}
\caption{The values of $c_0$, $c_1$, $c_2$ for which the one-link
 operators have no contact terms ($\lambda_\calO=0$).}
\label{ftildec2tab}
\vspace{0.5 cm}
\end{table} 

 For the one-link operators, we calculate the Green's functions in
 the limit $m^2 \ll p^2$, keeping terms up to first order in $m$. 
 Using the results of our $O(a)$ calculations which we have collected
 in the Appendix, we can derive the values of the renormalisation 
 constants and improvement coefficients which are given in the
 Tables~\ref{Z1link}-\ref{ftildec2tab}. For each operator there is 
 a particular value of the $c_i$s where the coefficient
 $\lambda_\calO$ vanishes, 
 and therefore there are no contact terms and even off-shell the operator 
 is simply renormalised by a multiplicative factor. These values are
 given in Table~\ref{ftildec2tab}. 

  Note that in the unpolarised case we can only determine 
 $c_1$ to $O(g^0)$ from our one-loop calculation.
 This is because $c_1$ is the coefficient 
 of an operator which vanishes at tree-level (because it 
 involves $[D_\nu , D_\lambda ]$).  However, we do still know
 the improved Green's function to $O(g^2)$.

 \section{Off-shell improvement for Ginsparg-Wilson fermions}  

 Like clover fermions, Ginsparg-Wilson fermions are free of $O(a)$ effects 
 on-shell. So it is instructive to see what happens when our 
 off-shell improvement conditions~(\ref{imp0_prop}) and~(\ref{imp0_3pt})
 are applied to Ginsparg-Wilson fermions~\cite{gw}.  
 
 The defining Ginsparg-Wilson relation is
 \begin{equation}
 D_{GW} \, \gamma_5 + \gamma_5 \, D_{GW}    
  = a  D_{GW} \,  \gamma_5  \, D_{GW} \,, 
 \label{GW_condition} 
 \end{equation} 
 where $D_{GW}$ is a Ginsparg-Wilson fermion matrix. 
 From this matrix we can (at least in
 principle) define a related matrix~\cite{chiu} 
 \begin{equation}
 K_{GW} \equiv \left( 1 - \frac{a}{2} D_{GW}\right)^{-1} D_{GW} . 
 \label{kdef} 
 \end{equation}
 The eigenvalues of $D_{GW}$ lie on a circle of radius $1/a$ 
 and centre $1/a$, while the eigenvalues of $K_{GW}$ lie on
 the imaginary axis.
 From eq.~(\ref{GW_condition}) and eq.~(\ref{kdef}) we find that 
 \begin{equation} 
 K_{GW} \, \gamma_5 + \gamma_5 \, K_{GW} =  0 .
 \end{equation}  
 The propagator which we would really like to know is the fermion
 propagator corresponding to $K_{GW}$. It should have the correct
 chiral properties, and be free from $O(a)$ discretisation 
 errors. However, we cannot work directly from  $K_{GW}$, because
 it is non-local. Therefore, we need to write down a formula for 
 the propagator we would get from $K_{GW}$, but expressed in terms 
 of $D_{GW}$.  
  This propagator will satisfy 
  chirality even at zero distance, so we expect it to
  be improved off-shell too. 

 Let us now add  mass to the problem in the same way as 
 it is added in the clover case, simply by adding 
 a mass term $m \psib \psi$ to the action, giving 
 \begin{equation}
  \psib \left[ D_{GW} + m \right] \psi 
 \label{this_action} 
 \end{equation}
 as the fermionic part of the action. Another way to 
 add mass effects would be to use the alternative action
 \begin{equation}
  \psib \left[ (1 - a m_\star/2 ) D_{GW} + m_\star \right] \psi ,
 \label{other_action} 
 \end{equation} 
 which has the advantage that there is no mass improvement  needed.
 That is the method we used in~\cite{gw}. Here 
 we have added the simple mass term, eq.(\ref{this_action}), 
 because we want to preserve the analogy with the clover action. 

 If we reexpress the unimproved massive propagator $(D_{GW} + m)^{-1}$
 in terms of $ K_{GW}$, we find 
 \begin{eqnarray}
 \lefteqn {\left\langle \frac{1}{D_{GW} + m} \right\rangle_A }\\
 &=& \frac{1}{(1 + \frac{a}{2} m )^2 } 
 \left\langle \frac{1}{ \left[ K_{GW} + m /(1 + \frac{a}{2} m ) \right] } 
 \right\rangle_A + \frac{a}{ 2 + a m} \delta(x-y) \nonumber. 
 \end{eqnarray} 
 Fourier transforming, we get 
 \begin{equation}
 S_{GW}(p,m) =  \frac{1}{(1 + \frac{a}{2} m )^2 } \; S_\star(p, m_\star)
 +  \frac{a}{ 2 + a m} \, , 
 \label{SGW_init} 
 \end{equation} 
 where 
 \begin{equation}
 m_\star \equiv  \frac{m} { 1 + \frac{a}{2} m }   
 \label{GW_mstar} 
 \end{equation} 
 and $S_\star$ is the Fourier transform of $\langle 
 ( K_{GW} + m_\star)^{-1}\rangle_A$. 
 Eqs.~(\ref{SGW_init}) and (\ref{GW_mstar}) are the analogues of 
 eqs.~(\ref{simprove}) and (\ref{mstar}) respectively,  
 remembering that terms of $O(a^2)$ are dropped in Sect.~\ref{propmethod}.  
 Solving eq.~(\ref{SGW_init}) for $S_\star$ we find
 \begin{equation}
 S_\star(p, m_\star) = (1 + \frac{a}{2} m )^2 
 \left[  S_{GW}(p,m) - \frac{a}{2 + a m }  \right] \,, 
 \label{GW_Sstar} 
 \end{equation} 
 which has the same form as  eq.~(\ref{imp0_prop}).
 Note that the only matrix we need to invert to calculate 
 this improved propagator is the matrix $(D_{GW} + m)$, which is
 well-defined, and local (in the sense that its elements decrease
 exponentially with separation). 
 Comparing these formulae with those in Sect.~\ref{propmethod}
 we see that 
 \begin{equation}
\lambda_\psi = 1 , \ \ \ \ b_\psi = -1 
 \ \ \ {\rm and} \ \ \ b_m = -\frac{1}{2} \;. 
 \end{equation}
 These results are
 independent of $g^2$. These all-order results coincide  with the 
 tree-level limit of the clover fermion result eq.~(\ref{propcoef}). 
 The values depend on the fact that in this paper
 we have added mass term to the Ginsparg-Wilson action in the same
 way as to the clover action. 
 
 Next we want to improve the Green's function corresponding to
 a flavour non-singlet operator $\calO \equiv \psib O \psi$,
 where $O$ includes Dirac structure and covariant derivatives. 
 We want our improved Green's function $G^\calO_\star(p,m_\star)$ to be 
 given by  
 \begin{equation}
 G^\calO_\star(p, m_\star) = \FT  
 \left\langle \frac{1}{K_{GW} + m_\star } O \frac{1}{K_{GW} + m_\star } 
 \right\rangle_A  \,, 
 \end{equation} 
 where ${\mathcal F}$ denotes the Fourier transform.
 However, we need to re-express it in a form that involves only
 $D_{GW}$, not $K_{GW}$. 
 This can be shown to be equivalent to the expression
 \begin{equation}
  G^\calO_\star(p,m_\star) =  
  ( 1 + \frac{a}{2} m)^2 
 \left[ G^{\calO^{imp}}_{GW}(p,m) -
 \frac{a}{2} \lambda_\calO C^\calO (p,m) \right] \,, 
 \label{GW3pt_1} 
 \end{equation}
 where
 \begin{eqnarray} 
  G^{\calO^{imp}}_{GW}(p,m)  &\equiv & \FT 
 \left\langle \frac{1}{D_{GW} + m} O^{imp} \frac{1}{D_{GW} + m} 
 \right\rangle_A \,,  \\
 O^{imp}  \equiv  & O & + \, a m c_0  O 
 - \frac{a}{2} c_1 ( D_{GW} O + O \, D_{GW} ) 
 + \frac{a^2}{4} D_{GW} O \, D_{GW} ,  
 \label{GWimpop1} \\ 
 C^\calO(p,m) &\equiv & \FT \left\langle O \frac{1}{D_{GW} + m} 
 + \frac{1}{D_{GW} + m} O \right\rangle_A , 
 \end{eqnarray}  
 with 
 \begin{eqnarray}
 c_0 &=& 1 - c_1  \,,\label{GW_impc} \\
 \lambda_\calO &=& 1 - c_1 . 
 \label{GW_implam} 
 \end{eqnarray} 
 Eq.~(\ref{GW3pt_1}) has the same form as eq.~(\ref{imp0_3pt})
 (up to terms of $O(a^2)$).
 A more general form of eq.~(\ref{GW3pt_1}) can be found in~\cite{gw}.  

 Comparing eqs.~(\ref{GW_impc}) and (\ref{GW_implam}) with 
 Tables \ref{bc1} and \ref{lamc1}, we see again that the all-orders
 Ginsparg-Wilson result is just the tree-level result for clover 
 fermions. A particular point to note is that the Ginsparg-Wilson
 improvement coefficients are the same for all operators, while 
 the clover action improvement coefficients are operator-dependent.  
 A further simplification in the Ginsparg-Wilson case is that there
 are no $\zeta$ coefficients needed, they are all zero. This means
 that the renormalisation constant $Z_\calO$ is independent of the 
 improvement coefficients $c_i$, (see eq.(\ref{defzeta})). 

 \section{Tadpole improvement}  
   
 \subsection{Analytic results}

 It is well known that many results from (naive) lattice perturbation theory 
 are in poor agreement with their counterparts determined from Monte
 Carlo calculations.  
 One main reason for this is the appearance of gluon tadpoles,
 which are typical 
 lattice artifacts. They make the bare coupling $g$ into a poor
 expansion parameter.
  Therefore, it was proposed~\cite{parisi,lepage}
  that the perturbative series should be rearranged
  in order to get rid of the numerically large tadpole contributions.  
 This rearrangement will be done 
 by using the variable $u_0$, derived from the measured value of the
 plaquette  
  \begin{equation} 
 u_0 = \left\langle \frac{1}{N_c} {\rm Tr} U_\Box \right\rangle^\frac{1}{4}.  
  \end{equation}   
 Its value depends on the coupling $g^2=6/\beta$ where it has been measured.
   
   There are two main steps involved in tadpole improvement.

 Firstly, we know~\cite{lepage} that in the mean field approximation the 
 $Z$ for an operator with $n_D$ derivatives is 
 \begin{equation}
 Z_\calO  \approx  u_0^{1 - n_D} \, ,
 \end{equation} 
 so it is reasonable to hope that a perturbative series for 
 $(Z_\calO u_0^{n_D-1})$ will converge more rapidly than a series for
 $Z_\calO$ itself. 
 Secondly, instead of writing our series in terms of the bare
 parameters, we reexpress it in terms of the tadpole improved, TI,
 parameters 
 \begin{eqnarray}
  g_\ti ^{2} &\equiv& g^2\;u_0^{-4} \, ,\nonumber \\
  \csw^{\ti}  &\equiv&  \csw \; u_0^3 \, , \nonumber \\
  c_i^\ti  &\equiv&  c_i \; u_0^n 
  \label{tadimp},  
 \end{eqnarray} 
 where  $n$ is the difference between the number of covariant
 derivatives in the higher dimensional operator multiplying $c_i$
 and the number of covariant derivatives in the operator to be improved 
 ($n$ is always 1 for our choice of improvement terms).
 The new coupling $g_\ti$ is called the ``boosted" coupling constant.
 Other choices of boosted coupling are possible, for example
 one could also  use the renormalised coupling constant at some scale
 $\approx 1/a^2 $. To carry out this rewriting of the series, we 
 simply replace every $g^2$ by $[g_\ti ^{2} u_0^4(g_\ti) ]$, where  
 $u_0(g_\ti)$ is the perturbative expansion for $u_0$:  
 \begin{equation}
 u_0(g_\ti) = 1 - \frac{g_\ti^2}{16 \pi^2} C_F \, \pi^2 + O(g_\ti^4)\,. 
 \end{equation}
 Formally, 
 this cannot change the all orders result, but it should improve the
 rate with which the series converges. The same procedure is followed
 with the improvement coefficients $\csw$ and $c_i$, for example 
 $\csw$ is to be replaced by $[\csw^\ti u_0^{-3}(g_\ti) ]$.      

   In this paper we will look at the tadpole improvement for operators
 with no anomalous dimension. The interplay between tadpole improvement
 and the renormalisation group, needed when considering operators 
 with an anomalous dimension, will be considered in a future paper. 
 The result of this procedure is rather simple for the one-loop 
 $Z$ factors. If the original $Z$ is given by 
 \begin{equation} 
 Z_\calO = 1 + \ggcf B_\calO(\csw , c_i)  +O(g^4) \, ,
 \end{equation} 
 then the tadpole improved $Z$ is given by 
 \begin{eqnarray}
 Z^\ti_\calO &=&  u_0^{1-n_D}\left[1+\frac{g_\ti ^{2}}{16\pi^2}C_F \, 
  \left( 
  B_\calO(\csw^\ti , c_i^\ti) 
 + (1 -n_D) \pi^2 \right) + O(g_\ti^4) \right] \nonumber \\
 &\equiv&  u_0^{1-n_D}\left[1+\frac{g_\ti ^{2}}{16\pi^2}C_F \, 
  B^\ti_\calO  +  O(g_\ti^4) \right] \, .
 \end{eqnarray} 

 For the $V$ and $A$ operators ($n_D=0$) in the $\MSB$ scheme we  
 get the following tadpole improved $B_\calO$ terms:
 \begin{eqnarray}  
 B^{\ti,\msb}_V  & = & -10.74819 + 4.74556\;\cswti +0.54317\; (\cswti)^2 
 \nonumber \\ & & 
 + \cti_1 \left(    9.78635\;   -  3.41640\; \cswti 
 - 0.88458\; (\cswti)^2 \right)
 ,\label{BOMSBltad} \\ 
 B^{\ti,\msb}_A  & = & -5.92668 - 0.24783\; \cswti +2.25137\; (\cswti)^2
  \nonumber  \\   & &
 + \cti_1 \left(   19.37225\; -   10.31673\; \cswti
  + 0.88458\; (\cswti)^2 \right) \; .
 \end{eqnarray}  
   
 Tadpole improvement is not just applicable to renormalisation
 factors -- it can also be used to give improved values for the
 improvement coefficients.  The improvement coefficients for the 
 fermion propagator, eq.~(\ref{propcoef}), become
 \begin{eqnarray} 
 \lambda_\psi &=& u_0^{-1}\left[ 1 +  
 \frac{g_\ti^2}{16 \pi^2} C_F\; (1.04124 - 1.85625 \; \alpha)  
 + O(g_\ti^4) \right] , \nonumber \\  
 b_\psi &=& - u_0^{-1}\left[ 1 + \ggcfti \; 6.52250  
 + O(g_\ti^4)  \right] , \nonumber \\  
 b_m &=& -\frac{1}{2} u_0^{-1}\left[ 1 + 
 \frac{g_\ti^2}{16 \pi^2} C_F \; 12.92445 + O(g_\ti^4) \right] .   
 \label{propcoef_tad} 
 \end{eqnarray} 
 
  An unfortunate ambiguity is that 
  there is of course considerable freedom in the choice of 
 boosted $g$. At one-loop none of the numerical coefficients are affected
 by this choice, so if one prefers another boosted $g$, all the
 formulae in this section can still be used, the only change
 is that every $g_\ti$ has to be replaced by the alternative 
 boosted $g$. 

\subsection{Comparison with non-perturbative results}

\begin{figure}[htbp]
  \begin{center}
    \epsfig{file=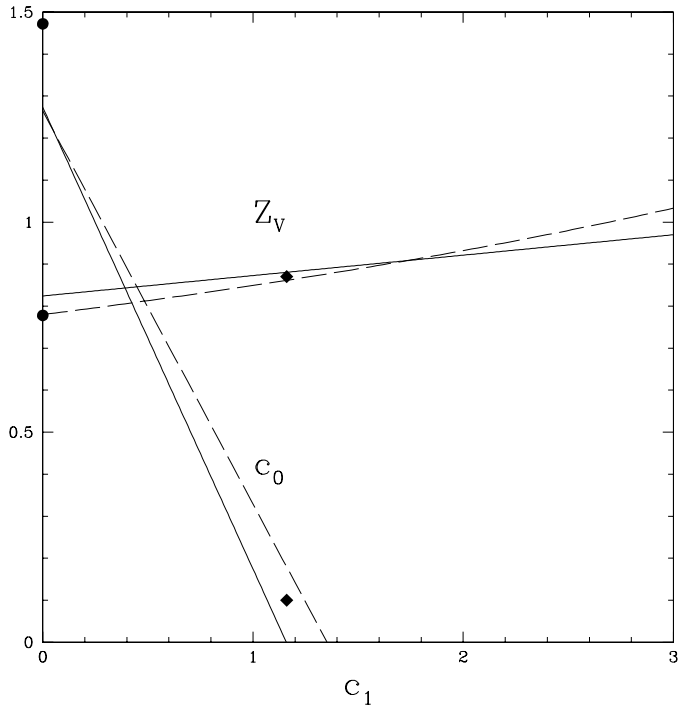,width=10.0cm,height=10.0cm}
\vspace*{0.3cm}
    \caption{The renormalisation constant $Z_V$ and the improvement
coefficient $c_0^V$ as a function of $\cmu$. The solid lines are the results
of tadpole improved perturbation theory. The dashed lines refer to the
non-perturbative results of~\cite{horslat97}, and the symbols mark the 
non-perturbative results of \cite{LSSW} (solid circle) and \cite{raklat97} 
(solid diamond).}
    \label{comp}
  \end{center}
\vspace{0.5 cm}
\end{figure}

To test the validity of tadpole improved perturbation theory, we will now 
compare our results with known non-perturbative results in the quenched 
theory. The local vector current is best suited for this purpose, because
the renormalisation constant $Z_V$ and improvement coefficient $c_0^V$ are
known non-perturbatively for a wide range of values of $\cmu$.

\begin{figure}[htbp]
  \begin{center}
    \epsfig{file=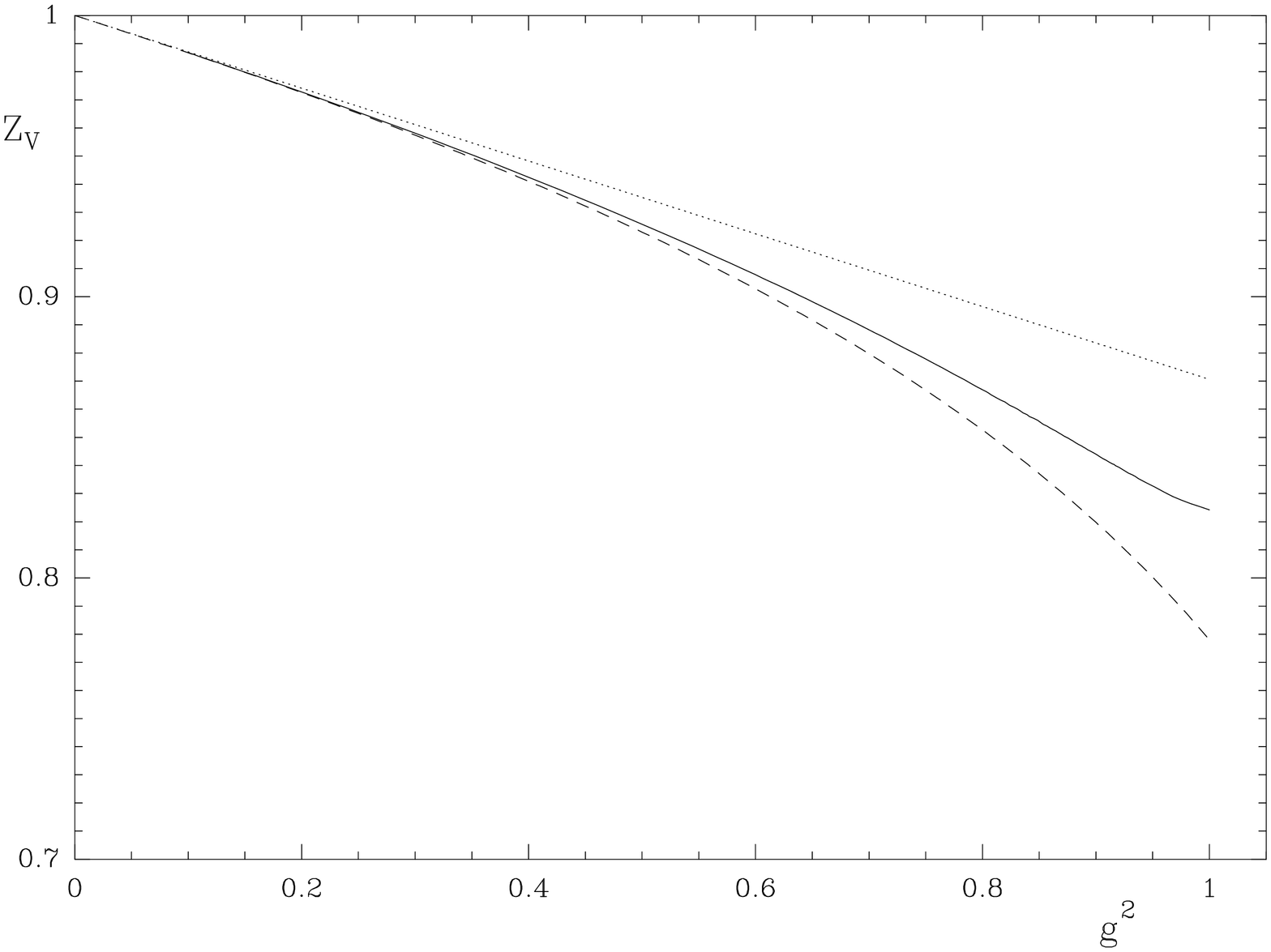, angle =270, width=13.0cm}
\vspace*{0.3cm}
    \caption{The renormalisation constant $Z_V$ as a function of $g^2$. 
 The solid line is the result of tadpole improved perturbation theory,
 while the dotted line shows the one-loop perturbative result with no
 improvement. The dashed line is the non-perturbative result of~\cite{LSSW}.}
    \label{g2comp}
  \end{center}
\vspace{0.5 cm}
\end{figure}

The comparison is done at $g^2 = 1$. At this value of the coupling one
finds non-perturbatively~\cite{LSSWW} $c_{SW} = 1.769$. We will use this 
number in both the perturbative formulae and the numerical calculations. 
For $u_0$ we obtain the value $0.8778$. We then get
\begin{eqnarray}
Z_V^\ti &=& 0.8242 + 0.0486  \cmu \,, \\
c_0^{\scriptscriptstyle{V},\ti} &=& 1.2733 - 1.0990 \cmu \;. 
\end{eqnarray}
In Fig.~\ref{comp} we show $Z_V^\ti$ and 
$c_0^{\scriptscriptstyle{V},\ti}$ as a function of
$\cmu$. We compare the results with the numbers of three independent
non-perturbative calculations. The first calculation~\cite{horslat97} uses
the nucleon matrix element of the local vector current to determine $Z_V$
and $c_0^V$. The second one is based on the Schr\"odinger 
functional~\cite{LSSW}, and the last calculation~\cite{raklat97} uses
 chiral Ward identities to improve the current and
 renormalisation following~\cite{mart1,nonpert}. 
 In the latter case we calculated only at the value of $\cmu$
  where $\lambda_V = 0$, as in Table~\ref{c1tild}. 
 It should be noted that the results still
have errors of $O(a^2)$, which can be as large as 10\%~\cite{qcdsf3b}, so 
that we cannot expect the results to agree completely.
For $Z_V$ we find good agreement between the tadpole improved perturbative
numbers and all the non-perturbative results. For $c_0^V$ our numbers agree
with~\cite{horslat97} and~\cite{raklat97}. The Schr\"odinger functional 
result, on the other hand, lies $\approx 10\%$ above the other numbers.
 (It is important to remember that different definitions of $Z$ may
 give results differing by $O(a^2)$, so both results could be 
 consistent.) 

 In Fig.~\ref{g2comp} we show the renormalisation constant $Z_V$ as
 a function of $g^2$.    
At smaller values of the coupling (higher values of $\beta$) the 
agreement between tadpole improved perturbation theory and non-perturbative 
results becomes even closer, as one might expect. In those cases where we 
could check this, we found the discrepancy to reduce to $\approx 4\%$ at 
$\beta = 6.4$. 
 Thus we may say that the non-perturbative results agree with 
those of tadpole improved perturbation theory within the expected $O(a^2)$ 
 and $O(g^4)$ uncertainties.

 \section{Conclusion}

 In this paper we have presented extensive one-loop perturbative
 calculations of lattice Green's functions, in which we have kept all
 $O(a)$ terms. This allows us to investigate operator improvement,
 firstly to see what sort of improvement terms are needed, and secondly
 to calculate values of the improvement coefficients. 

  We find that we can produce off-shell $O(a)$ improved Green's functions, 
 to all orders in the Ginsparg-Wilson case, and at least to $O(g^2)$
 in the clover case.  In our one-loop calculations we find that we only 
 need gauge-invariant improvement terms. No extra improvement terms 
 associated with BRST symmetry are required at this level. 

  Off-shell improvement doesn't
 mean that there are no contact terms. As long as we know the
 form of the contact terms, we can remove them by using the 
 improvement coefficients $\lambda$. 
  Contact terms, responsible for the off-shellness of the 
 propagators and Green's functions, can be removed using a 
 well-determined procedure.
 There are always particular values of the improvement coefficients
 for which the contact terms vanish, so that
 one still has a multiplicative renormalisation.

  In the Ginsparg-Wilson case improvement is particularly simple, 
 because the improvement coefficients are universal, they do not
 depend on the operator considered, the coupling constant, or even on
 which theory we are simulating (we assume that the bosonic
 sector has no $O(a)$ discretisation errors).  This is not so
 in the clover case, the coefficients depend on the coupling, 
 and are different for each operator. 

   We have the tadpole improved one-loop values for $Z$
 factors calculated at arbitrary  $\csw$, and for 
 improvement coefficients calculated at $\csw = 1 + O(g^2)$, 
 which is the only place where $O(a)$ improvement is possible.
  Numerical test cases show that tadpole improvement
 works well down to $\beta \approx 6.0$ for operators with no 
 anomalous dimension. We are investigating tadpole improvement in 
 the case of operators with an anomalous dimension. 

 \section{Acknowledgements}

  S.C. is supported in part by the U. S. Department of Energy (DOE) 
 under cooperative research agreement DE-FC02-94ER40818.
 
 Support by the Deutsche Forschungsgemeinschaft and by the BMBF is
 also gratefully acknowledged. 


 \appendix

 \section{Appendix}    

 In this Appendix we give the perturbative expressions for the 
 amputated three-point functions 
 (vertex functions) for all the operators we have considered
 (apart from the scalar density, which is given in Sect.~\ref{point_sect}
 of the main text), 
 calculated to $O(a)$, and the values of their improvement coefficients. 
 In order to make transparent the transformation of these numbers   
 into the popular $\MSB$ scheme the corresponding 
 continuum quantities are also given.  

 In order to shorten the expressions for the Green's functions  
 we will  use the functions $T$ and $L$ defined in eq.~(\ref{TLdef}). 

 \subsection{Pseudoscalar Vertex} 

 The pseudoscalar operator is simpler than the other operators  
 because there is no $c_1$ improvement term possible (and also 
 none needed) for the  
 forward three-point function, because ${\not{\hspace{-0.1cm}{D}}}$ 
 and $\gamma_5$ anti-commute.  
 The one-loop expression for this vertex up to $O(a)$ is:
 \begin{eqnarray}
 \Lambda^P(p,m; c_0=0) &=& \gamma_5 + \gamma_5 \, \ggcf
    \Big[ -1.42500   + 2 \,\alpha  + 3.43328 \,{\csw^2}
 \nonumber \\ &&
   -  ( 3 + \alpha  )  \, L(a p, a m)  -  ( 3  + \alpha ) \Tterm
 \Big] \nonumber \\
 & & +  a\,m \, \gamma_5 \, \ggcf \Big[ 3.82788
      - 0.14375  \,\alpha - 2.49670 \,{\csw^2}
 \nonumber \\ &&
    +  ( 3 + \alpha  )  L(a p, a m) + 2 \,( 3  + \alpha )   \Tterm
  \Big] .
\end{eqnarray}
In the $\MSB$ scheme we have 
\begin{eqnarray}
 \Lambda^P_\msb(p,m_\msb)  =  \gamma_5 + \gamma_5 \, \ggcf
       \Big[  && \, 4 + 2\,\alpha
   - ( 3 + \alpha  ) \, \ln \left( \frac{p^2 + m_\msb^2}{\mu^2} \right) 
   \nonumber \\
 &&  -  ( 3 + \alpha )  T\left(p^2/m_\msb^2 \right)
  \Big] .
 \end{eqnarray}

 The MOM scheme renormalisation factor is defined by 
 \begin{equation}
  Z^{\rm MOM}_2(M^2)  Z^{\rm MOM}_P(M^2) = 
 \frac {4 N_c} { {\rm Tr}\left[ \gamma_5 \Lambda^P(p,m;0) \right] }
 \end{equation} 
 at $p^2 = M^2$.  The result is 
\begin{eqnarray}
 Z^{\rm MOM}_P(M^2)  = 1 &+& \ggcf \Big[ 
 -15.21941  - \alpha + 2.24887 \csw - 2.03602 \csw^2 
 \nonumber \\ && 
   + 3 \starLapam +  (3 + \alpha) \starTterm  
 \nonumber \\ && 
 - \alpha \starminT \Big] +O(a) .
\end{eqnarray}
 As well as the lack of a derivative improvement term, another
 special feature of the pseudoscalar operator is that the 
 improvement terms $a m \Lambda^P$ and $a \{ S^{-1} , \gamma_5 \}$, 
 which appear in eq.~(\ref{imp0_Lambda_Gam}), 
 have the same functional form to this order, so there is no
 natural way of determining $c_0^P$ and $\lambda_P$ separately. We
 choose to improve the operator by setting  $\lambda_P = 0$, and making 
 all the improvement through the mass-dependent term. 
 Then the  improvement coefficients are (at $\csw = 1$) 
 \begin{eqnarray}
 c_0^P &=& - \ggcf 5.21443 \,, \nonumber \\
 \lambda_P &\equiv& 0 . 
 \end{eqnarray}

\subsection{Local Vector} 

 The one-loop expression for the local vector vertex up to $O(a)$ is:
 \begin{eqnarray} 
 \lefteqn{ 
 \Lambda^V_\mu(p,m;0,\cmu) =  \gamma_\mu
 - a\,{\rm i} p_\mu\,\cmu }
 \nonumber \\
 &+&  \gamma_\mu \ggcf \Big[ 3.97338 + \alpha 
  - 2.49670 \csw    + 0.85410 \csw^2    - \cmu ( 9.78635
 \nonumber \\ & & 
  -  3.41640 \csw  -  0.88458 \,{\csw^2} )   
    - \alpha  \,L(a p, a m)  - \alpha \minT 
  \Big] \nonumber \\ 
  & +& \frac{\pslash \, p_\mu}{p^2}\, \ggcf \,   
      \Big[-2 \,\alpha +4 \,\alpha\frac{m^2}{p^2} 
 \left( 1 - \Tterm \right) \Big] \nonumber \\ 
&+&  {\rm i} \frac{m p_\mu}{p^2}\, \ggcf \,   
      \Big[ 2 \,(3+\alpha) \left( 1- \Tterm \right)\Big] \nonumber\\ 
&+&   a {\rm i} p_\mu \ggcf \Big[-8.66505
 + 2.85625 \,\alpha +9.52789 \csw -0.39053 \csw^2 
 \nonumber \\ && 
 +\cmu (18.59361   - \alpha - 2.24887 \csw - 0.16098 \csw^2 ) 
 \nonumber \\ && 
   -(3 -\alpha -\,\cmu \,\alpha - 3 \csw) L(a p,a m)  
 + 2 (3 + \alpha ) \frac {m^2}{m^2 + p^2} 
 \nonumber \\ & & 
 - (18 + 8\, \alpha - 6\, \cmu - 3 \, \alpha \cmu - 3\csw ) \minT 
\Big] \nonumber \\ 
&+&    a m \gamma_\mu\, \ggcf \,   
      \Big[-7.64168 +0.85625 \,\alpha +7.74287 \csw - 1.38589 \csw^2 
 \nonumber \\ && 
 + \cmu ( 13.96523  - \alpha   - 5.54358 \csw - 0.36162 \csw^2 ) 
 \nonumber \\ && 
 - \half \left( 3 -6\,\cmu -2\,\alpha-3\csw \right)\,L(a p,a m)
 \nonumber \\ && 
 - (  3 - 3\csw - 3\,\cmu - \alpha \, \cmu ) \Tterm 
 \nonumber \\&& 
 + \half \left(3 +6 \, \alpha - 3 \csw 
 - 2\, \alpha \, \cmu \right) \minT  
\Big] \\ 
&+&   a \frac{m \pslash \, p_\mu}{p^2} \ggcf \Big[3+4 \,\alpha
 -2 \,\cmu \,\alpha + 3 \csw
 - 2 ( \alpha + 3\csw ) \Tterm  
 \nonumber \\&& 
 +4 \,\alpha \frac{m^2}{m^2+p^2}
 - \Big( 6 + 12\,\alpha -4 \,\cmu \,\alpha
 - 6 \csw \Big)\minT 
 \Big] \nonumber .
\end{eqnarray} 
In the $\MSB$ scheme we have 
 \begin{eqnarray} 
 \Lambda^V_\msb(p,m_\msb) &=&  \gamma_\mu  
  +   \gamma_\mu \ggcf \left[ \alpha 
    - \alpha  
 \, \ln \left( \frac{p^2 + m_\msb^2}{\mu^2} \right) \right.
 \nonumber \\ && 
  \qquad \qquad  \left. - \alpha 
 \, \frac{m_\msb^2}{p^2} \left( 1 - T \left(p^2/m_\msb^2 \right)\right) 
  \right] \nonumber \\ 
 & +& \frac{\pslash \, p_\mu}{p^2}\, \ggcf \,   
      \Big[-2 \,\alpha + 4 \,\alpha
 \, \frac{m_\msb^2}{p^2} \left( 1 - T \left(p^2/m_\msb^2 \right)\right) 
 \Big] \nonumber \\ 
&+&  {\rm i} \frac{m p_\mu}{p^2}\, \ggcf \,   
      \Big[ 2 \,(3+\alpha) 
 \left( 1 - T \left(p^2/m_\msb^2 \right)\right)
 \Big] .
\end{eqnarray} 
  We consider two ways to define $Z^{\rm MOM}$ for the vector, 
\begin{eqnarray} 
 \frac{1}{3} \sum_{\mu \nu}
 \left( \delta_{\mu\nu} -\frac{p_\mu p_\nu} {p^2} \right)
 \frac{1}{4 N_c} {\rm Tr} \left[ \gamma_\mu \Lambda^\star_\nu \right] 
 &\equiv& \frac{1}{ Z_2^{\rm MOM} Z^{\rm MOM}_{V_{\rm trans}} },
 \label{transdef}\\
 \sum_\mu  \frac{p_\mu}{p^2}
 \frac{1}{4 N_c} {\rm Tr} \left[ \pslash \Lambda^\star_\mu \right]
 &\equiv& \frac{1}{Z_2^{\rm MOM} Z^{\rm MOM}_{V_{\rm long}} }, 
 \label{longdef} 
\end{eqnarray} 
so that
\begin{eqnarray} 
 Z^{\rm MOM}_{ V_{\rm trans}}(M^2; 0) 
 \mkern 40mu  &&  \\
 = 1 + \ggcf \big[ 
 &-&20.61780 + 4.74556 \csw + 0.54317 \csw^2 \big] +O(a), 
  \nonumber \\  
 Z^{\rm MOM}_{ V_{\rm long}}(M^2; 0) = 1 &+& \ggcf \Big[ 
 - 20.61780 + 2 \, \alpha + 4.74556 \csw  
 \\ && 
 + 0.54317 \csw^2 - 4 \alpha \starminT \Big] +O(a) . \nonumber 
\end{eqnarray}

\subsection{Conserved Vector}

 The one-loop expression for the conserved vector vertex up to $O(a)$ is:
  \begin{eqnarray} 
  \Lambda^J_\mu(p,m) &=& 
 \gamma_\mu - a\,{\rm i}\, p_\mu \nonumber \\
  &+&  \gamma_\mu \, \ggcf \Big[
    - 16.64441 + \alpha + 2.24887\csw + 1.39727\csw^2
 \nonumber \\ & & 
    - \alpha  \,L(a p, a m) 
  - \alpha \minT \Big] \nonumber \\ 
  &+&    \frac{\pslash \, p_\mu}{p^2}\, \ggcf \,   
      \Big[-2 \,\alpha +4 \,\alpha\frac{m^2}{p^2} 
  \left( 1 - \Tterm \right) \Big] \nonumber \\ 
  &+&   {\rm i} \frac{m p_\mu}{p^2}\, \ggcf \,   
      \Big[ 2 \,(3+\alpha) \left( 1- \Tterm \right)\Big] \nonumber\\ 
  &+&   a {\rm i} p_\mu \ggcf \Big[
  11.27782 + 1.85625  \, \alpha  + 3.97134 \csw  - 0.16345 \csw^2
 \nonumber  \\ && 
   -(3 - 2 \, \alpha - 3 \csw) L(a p,a m)  
  + 2 (3 + \alpha ) \frac {m^2}{m^2 + p^2} 
  \nonumber \\ & & 
  - (12 + 5\, \alpha - 3\csw ) \minT 
  \Big] \nonumber \\ 
  &+&    a m \gamma_\mu\, \ggcf \,   
      \Big[
   6.34664 - 0.14375\,\alpha + 1.48503\csw - 1.28605\csw^2
 \nonumber \\ && 
  +\half \left( 3  +2\,\alpha +3\csw \right)\,L(a p,a m)
  + ( 3\csw + \alpha ) \Tterm 
  \nonumber \\&& 
  + \half \left(3 +4 \, \alpha - 3 \csw \right) \minT  
  \Big] \\ 
   &+&   a \frac{m \pslash \, p_\mu}{p^2} \ggcf \Big[3+2 \,\alpha
   + 3 \csw
   - 2 ( \alpha + 3\csw ) \Tterm  
 \nonumber  \\&& 
   +4 \,\alpha \frac{m^2}{m^2+p^2}
   - \Big( 6 + 8\,\alpha - 6 \csw \Big)\minT 
   \Big] \nonumber .
  \end{eqnarray} 
In the $\MSB$ scheme we have 
 \begin{eqnarray} 
 \Lambda^J_\msb(p,m_\msb) &=&  \gamma_\mu  
  +   \gamma_\mu \ggcf \left[ \alpha 
    - \alpha  
 \, \ln \left( \frac{p^2 + m_\msb^2}{\mu^2} \right) \right.
 \nonumber \\ && \qquad \qquad \left.
   - \alpha 
 \, \frac{m_\msb^2}{p^2} \left( 1 - T \left(p^2/m_\msb^2 \right)\right) 
  \right] \nonumber \\ 
  & +& \frac{\pslash \, p_\mu}{p^2}\, \ggcf \,   
      \Big[-2 \,\alpha + 4 \,\alpha
 \, \frac{m_\msb^2}{p^2} \left( 1 - T \left(p^2/m_\msb^2 \right)\right) 
 \Big] \nonumber \\ 
&+&  {\rm i} \frac{m p_\mu}{p^2}\, \ggcf \,   
      \Big[ 2 \,(3+\alpha) 
 \left( 1 - T \left(p^2/m_\msb^2 \right)\right)
 \Big] .
\end{eqnarray} 
 We can define longitudinal and transverse $Z^{\rm MOM}$ 
 according to eqs.~(\ref{transdef}) and~(\ref{longdef}), 
 giving us 
\begin{eqnarray}
 Z^{\rm MOM}_{J_{\rm trans}}(M^2) &=& 1 + O(a) 
  \\  
 Z^{\rm MOM}_{J_{\rm long}}(M^2) &=& 1  +  \ggcf \left[ 
  2 \, \alpha - 4 \alpha \starminT \right] + O(a) .
\end{eqnarray}
  The conserved vector current satisfies eq.~(\ref{imp0_3pt}) 
 without any improvement terms, i.e. 
\begin{eqnarray}
c_0^J &=& 0 \,, \nonumber \\
\lambda_J &=& 0 . 
\end{eqnarray}
This is as it should be, the conserved current is already 
 improved for forward matrix elements, and no further 
 improvement terms are needed.

\subsection{Axial Vector}

 The one-loop expression for the axial vector vertex up to $O(a)$ is:
 \begin{eqnarray}
 \lefteqn{ 
 \Lambda^A_\mu (p,m;0,\cax) = \gamma_\mu \gamma_5 - a \pvstruc \cax 
 } \nonumber \\ 
  &+&   \gamma_\mu \gamma_5 \ggcf 
  \, \Big[ -0.84813 + \alpha + 2.49670  \csw - 0.85410  \csw^2   
 \nonumber \\ &&
     - \cax \, ( 19.37225  
  - 10.31673 \csw + 0.88458 \csw^2) - \alpha \, L(a p, a m) 
 \nonumber  \\ && 
       - 2 (1 + \alpha) \, \Tterm 
   -( 2 - \alpha ) \minT \Big] \nonumber \\ 
   &+&  \frac {p_\mu \, \pslash }{p^2} \gamma_5 
   \, \ggcf \left[ -2 \, \alpha
   - 4 \, ( 1 - \alpha) \, \Tterm 
 \right. \nonumber \\ && \left.
  + 4 (2 -\alpha )\, \minT 
   \right]  \nonumber \\
   &+& \pvstruc \frac{m}{p^2} \, \ggcf 
  \left[ 2 \, ( 1 - \alpha) \,\left( 1 - \Tterm \right) \right] \nonumber\\
  &+& a \pvstruc \ggcf \, \Big[ 1.34275  + 0.85625  \, \alpha
  - 1.71809  \csw
  \nonumber \\ && 
  + 0.13018  \csw^2  
  + \cax \, (13.96522  - \alpha + 0.54301  \csw  + 0.05366  \csw^2) 
  \nonumber \\ && 
  + \alpha \, (1 +  \cax) \, L(a p, a m)   
 \nonumber \\ && 
       + 2 (\alpha + \csw) \, \Tterm + 2 \, ( 1 - \alpha) \, \mmpole   
 \nonumber \\ && 
       - ( 1 - 4 \, \alpha - 2 \cax 
 + \alpha \cax + 2 \, \csw) \, \minT 
  \Big]  \nonumber \\    
  &+&  a m  \gamma_\mu \gamma_5 \ggcf 
   \Big[  7.47851  - 1.14375  \, \alpha - 
         8.74287 \csw + 1.38589 \csw^2   
  \nonumber \\ && 
      +  \cax \, (10.92421 
   - \alpha - 4.44321  \csw + 0.36162  \csw^2)   
 \nonumber \\ && 
       + \half \left(6 \, \cax + 3 + 2\,\alpha - 3 \csw
  \right) \, L(a p, a m)   
 \nonumber \\ && 
       + (3 + 6\, \alpha + (1 - \alpha) \, \cax - \csw) \, \Tterm 
   - 4 \, \mmpole   
\nonumber  \\ && 
       + \half \left( 13 -6 \, \alpha 
 - 2\,(2 - \alpha) \, \cax  - \csw \right) \, \minT 
  \Big] \nonumber \\
  &+& a m \frac {p_\mu \, \pslash }{p^2} \gamma_5 \ggcf 
   \, \Big[ -3 + 4 \, \alpha - 2 \, \alpha \, \cax + \csw   
         + 4 \, \mmpole   
 \nonumber  \\ && 
 + 2 \, (6 - 5 \, \alpha - 2 \, (1 - \alpha) \, \cax - \csw ) \, \Tterm 
  \nonumber \\ && 
  - 2 \, ( 13 - 6 \, \alpha - 2 \, (2 - \alpha) \, \cax - \csw ) \, \minT   
   \Big] .
  \end{eqnarray}

In the $\MSB$ scheme we have 
\begin{eqnarray}
 \lefteqn{ 
 \Lambda^A_\msb(p,m_\msb) = \gamma_\mu \gamma_5  
  +    \gamma_\mu \gamma_5 \ggcf 
 \, \Big[ 
 \alpha - \alpha \, \mslog }
 \nonumber \\ && 
      - 2 \, (1 + \alpha) \, \msTterm  
 -(2 -\alpha )\, \msminT \Big]  \nonumber \\ 
  &+&  \frac {p_\mu \, \pslash }{p^2} \gamma_5 
  \, \ggcf \Big[ -2 \, \alpha
  - 4 \, ( 1 - \alpha) \, \msTterm
  \nonumber \\ && \qquad \qquad 
 + 4 \,(2 -\alpha)\, \msminT 
 \Big]  \nonumber \\
  &+& \pvstruc \frac{m_\msb}{p^2} \, \ggcf 
 \Big[ 2 \, ( 1 - \alpha) \,\left( 1 - \msTterm \right) \Big] .
 \end{eqnarray} 
  Again we can define both transverse and longitudinal $Z^{\rm MOM}$, 
\begin{eqnarray} 
 \frac{1}{3} \sum_{\mu \nu}
 \left( \delta_{\mu\nu} -\frac{p_\mu p_\nu} {p^2} \right)
 \frac{1}{4 N_c} {\rm Tr}
 \left[\gamma_5 \gamma_\mu \Lambda^\star_\nu \right] 
 &\equiv& \frac{1}{ Z_2^{\rm MOM} Z^{\rm MOM}_{A_{\rm trans}} },
 \\
 \sum_\mu  \frac{p_\mu}{p^2}
 \frac{1}{4 N_c} {\rm Tr} \left[ \gamma_5 \pslash \Lambda^\star_\mu \right]
 &\equiv& \frac{1}{Z_2^{\rm MOM} Z^{\rm MOM}_{A_{\rm long}} }. 
\end{eqnarray} 
 This gives the MOM scheme renormalisation factors
\begin{eqnarray}
 Z^{\rm MOM}_{A_{\rm trans}}(M^2;0) = 1 &+&  \ggcf \Big[ 
 -15.79628  - 0.24783 \csw + 2.25137 \csw^2
 \nonumber \\ && 
  + 2 \, (1 +\alpha) \starTterm 
 \nonumber \\ && 
 + 2 \, (1 -\alpha) \starminT \Big] + O(a) 
  , \\  
 Z^{\rm MOM}_{A_{\rm long}}(M^2;0) = 1 &+& \ggcf \Big[ 
 -15.79628  - 0.24783 \csw + 2.25137 \csw^2
 \nonumber \\ && 
   + 2 \, \alpha + 2 \, (3 -\alpha) \starTterm
 \nonumber \\ && 
 - 2 \, ( 3 -\alpha) \starminT \Big] +O(a) . 
\end{eqnarray}

\subsection{Tensor vertex} 

 The one-loop expression for the tensor vertex up to $O(a)$ is:
\begin{eqnarray}
 \lefteqn{ \Lambda^H_{\mu \nu} (p, m; 0, \ct) 
 = \sigma_{\mu\nu}\gamma_5 - a  \ct \tenstruc }
 \nonumber \\ &+& 
  \sigma_{\mu \nu} \gamma_5 
  \ggcf \Big[ 4.16568 - 1.66446 \csw - 0.57503  \csw^2  
 \nonumber \\ &&
      - \ct ( 16.24376  
 - 6.85531 \csw - 0.58972 \csw^2) + (1 - \alpha) L(a p, a m)  
 \nonumber \\ &&
     - ( 1 - \alpha) \Tterm  + 2 ( 1 - \alpha) \minT \Big] 
 \nonumber \\ &+& {\rm i}\, 
 \frac{ \pslash}{p^2} (\gamma_\nu p_\mu - \gamma_\mu p_\nu) \gamma_5 
  \ggcf \Big[ 2 ( 1 - \alpha) \Tterm 
 \nonumber \\
 && - 4  ( 1 - \alpha) \minT \Big]
 \nonumber \\ &+& 
  \tenstruc \frac{m}{p^2} \ggcf \Big[ 4  \left(1 - \Tterm \right) \Big]  
 \nonumber \\  &+& 
 a \tenstruc \ggcf 
    \Big[ -3.66115 + 1.85625 \alpha + 0.96286 \csw  
 \nonumber \\ && 
  + 0.42868 \csw^2 + \ct (16.27942  - \alpha + 0.26479 \csw - 0.26155 \csw^2) 
 \nonumber \\ && 
 + \half \left(2\,\alpha \ct - 3 + 2\,\alpha + \csw \right) L(a p, a m)  
         + (\alpha - \csw) \Tterm 
 \nonumber \\ && 
  + \half \left(2\,(4 + \alpha) \ct 
     - 19 - 4 \,\alpha + 3 \csw \right) \minT 
 \nonumber \\ && 
       + 4 \mmpole  
  \Big]
 \nonumber \\ &+& 
    a m \sigma_{\mu \nu} \gamma_5  \ggcf  
        \Big[ -7.42480  + 1.85625  \alpha +
              5.16191 \csw - 0.09170  \csw^2  
 \nonumber \\ && 
      + \ct (17.60663 - 2 \alpha - 7.02465 \csw - 0.24108 \csw^2)  
 \nonumber \\ && 
            - ( 2 - \alpha - 4 \ct - \csw)  L(a p, a m)  
 \nonumber \\ && 
         + 2 \, ( (1 + \alpha) \ct -  \alpha + \csw)) \Tterm 
 \nonumber \\ && 
           -  ( 5 - 6 \alpha - 2 ( 1 - \alpha) \ct + \csw) \minT  
 \Big]
 \nonumber \\ &+& {\rm i} a m 
 \frac{ \pslash}{p^2} (\gamma_\nu p_\mu - \gamma_\mu p_\nu) \gamma_5 
  \ggcf  \Big[3 + \csw 
 \nonumber \\ && 
   + 2 \left( ( 1 - \alpha) \ct - ( 3 - 2 \alpha + \csw) \right) \Tterm 
 - 2 ( 1 - \alpha) \mmpole  
 \nonumber \\ && 
 + 2 \left(-2 ( 1 - \alpha) \ct +  5 - 6 \alpha + \csw \right) \minT 
 \Big] .
 \end{eqnarray}
In the $\MSB$ scheme we have 
 \begin{eqnarray}
 \lefteqn{ 
 \Lambda^H_\msb (p, m_\msb ) = \sigma_{\mu \nu} \gamma_5 } 
 \nonumber \\ && + 
 \sigma_{\mu \nu} \gamma_5 \ggcf (1 -\alpha) 
 \Big[ \mslog - \msTterm 
 \nonumber \\ && 
 \qquad \qquad + 2 \msminT \Big] \nonumber \\ && 
 + {\rm i}\frac{ \pslash}{p^2} (\gamma_\nu p_\mu - \gamma_\mu p_\nu) \gamma_5
 \ggcf (1 -\alpha ) \Big[ 2 \msTterm 
 \nonumber \\ &&
 \qquad \qquad \qquad  - 4 \msminT \Big] \nonumber \\ && 
 + \tenstruc \ggcf  \frac{ m_\msb}{p^2} 4 \left[ 1 - \msTterm \right] . 
 \end{eqnarray}

 We define $Z^{\rm MOM}$ by 
 \begin{equation} 
 \frac{1}{12} \sum_{\mu \nu} \frac{1}{4 N_c} 
 {\rm Tr} \left[ \gamma_5 \sigma_{\mu \nu} \Lambda^\star_{\mu \nu} \right] = 
 \frac{1}{Z_2^{\rm MOM} Z_H^{\rm MOM} }  
 \end{equation} 
 at the scale $p^2 = M^2$, which gives
\begin{eqnarray}
 \lefteqn{ 
 Z^{\rm MOM}_H(M^2;0) = 1 + \ggcf \Big[
 -20.81009  + 3.91333 \csw +  1.97230 \csw^2}
 \nonumber \\ && \qquad 
   +\, \alpha - \starLapam - \alpha \starminT \Big] +O(a) .
\end{eqnarray}

 \subsection{ First unpolarised moment, off-diagonal ($\mu \ne \nu$),
 symmetrised \label{sectunpold} }

 For the one-link operators, we calculate the Green's functions in
 the limit $m^2 \ll p^2$, keeping terms up to first order in $m$. 
 This is sufficient to calculate the improvement coefficients. 

 Our improved operator for the first unpolarised moment in the 
 $\tau_3^{(6)}$ representation of the hypercubic group is 
  \begin{eqnarray}
 \calO_{\mu\nu}^{\tau_3^{(6)}} & = &
 \left(1+ a\, m\, c_0 \right) \frac{1}{2} \; \psib \frac{1}{2} 
 \left(\gamma_\mu \Dlr_\nu + \gamma_\nu \Dlr_\mu \right) \psi
 \nonumber \\ &&
  + \frac{1}{8}\,a\,{\rm i} c_1  \sum_\lambda \psib
 \frac{1}{2} \left(
 \sigma_{\mu\lambda}\left[\Dlr_\nu,\Dlr_\lambda\right]
 +\sigma_{\nu\lambda}\left[\Dlr_\mu,\Dlr_\lambda\right]
  \right) \psi \nonumber \\
 & & -\frac{1}{8} \,a \,c_2
  \psib\left\{\Dlr_\mu,\Dlr_\nu\right\}\psi \, . 
 \end{eqnarray}
 The result for the amputated operator Green's function is:
 \begin{eqnarray}
 \lefteqn{ \Lambda^{\tau_3^{(6)}}(p,m;0,c_1,c_2)  = 
 {\rm i} \frac{1}{2} ( \gamma_\mu p_\nu +  \gamma_\nu p_\mu )
  + c_2 a p_\mu p_\nu }
 \nonumber \\
 &+& {\rm i} \frac{1}{2} ( \gamma_\mu p_\nu +  \gamma_\nu p_\mu )
 \ggcf \Big[ 
 (\frac{8}{3} -\alpha ) \ln a^2 p^2 
 -18.80927 + 3.79201 \alpha 
 \nonumber \\ && 
 - 1.62411 \csw + 0.71900 \csw^2 
 + c_1  (-4.27417 + 1.08793 \csw)
 \nonumber \\ && 
 + c_2 ( -9.40584 + 4.60327 \csw + 0.46669 \csw^2 ) \Big] 
 \nonumber \\  
 & + & {\rm i} \frac{ p_\mu p_\nu}{p^2} \pslash \ggcf 
 \Big[ - \frac{1}{3} - \alpha \Big] 
  - m \frac{ p_\mu p_\nu}{p^2} \ggcf 4 
 \nonumber \\  
 & + & a p_\mu p_\nu \ggcf \Big[ 
 (3 -\alpha -\frac{4}{3} \csw ) \ln a^2 p^2 
  -2.80639 + 1.43576 \alpha
 \nonumber \\ && 
  - 0.49196 \csw + 0.10443 \csw^2
 \nonumber \\ && 
 + c_1 \left( - \frac{1}{3} \ln a^2 p^2 + 0.86075 - 0.30348 \csw \right)
 \nonumber \\ && 
 + c_2 \left( (\frac{4}{3} -\alpha )  \ln a^2 p^2 
 -33.31690 + 4.29201 \alpha 
 \right. \nonumber \\ && \left.
 + 1.16439 \csw + 0.04840 \csw^2 \right) \Big] 
 \nonumber \\  
 &+& {\rm i} a m \frac{1}{2} ( \gamma_\mu p_\nu +  \gamma_\nu p_\mu )
 \ggcf \Big[
 \left( -\frac{13}{6} + \alpha + \frac{1}{2} \csw \right) \ln a^2 p^2
 \nonumber \\ &&  
 + 1.76784 - 2.93576 \alpha
 + 2.56080 \csw - 0.92094 \csw^2 
 \nonumber \\ &&  
  + c_1 \left(  \ln a^2 p^2 -2.06261 - 0.62935 \csw \right) 
 \nonumber \\ &&  
 + c_2 \left( \frac{11}{3} \ln a^2 p^2 
 + 0.56785 - \alpha - 4.93253 \csw - 0.19761 \csw^2 \right) 
 \Big] \nonumber \\   
 & + & a m  {\rm i} \frac{ p_\mu p_\nu}{p^2} \pslash \ggcf
 \Big[ \frac{7}{3} + 2 \alpha + \csw 
 + c_2 ( \frac{2}{3} - 2 \alpha ) \Big] .
 \label{unfirstd}
 \end{eqnarray}
 The contact Green's function is given by 
 \begin{equation}
 C^\calO(p,m) = \Delta^\calO (p,m) S(p,m)        
 + S(p,m) {\overline \Delta}^\calO (p,m)  \,, 
 \end{equation} 
 where 
 \begin{eqnarray}
 \lefteqn{ \Delta^{\tau_3^{(6)}} (p,m) = 
 {\rm i} \frac{1}{2} ( \gamma_\mu p_\nu +  \gamma_\nu p_\mu ) }
 \nonumber \\
 && + {\rm i} \frac{1}{2} ( \gamma_\mu p_\nu +  \gamma_\nu p_\mu )
 \ggcf ( \ln a^2 p^2 -14.27168 - \frac{1}{2} \alpha + 0.19740 \csw ) 
 \nonumber \\ &&
 + \frac{m}{p^2}  \frac{1}{2} ( \gamma_\mu p_\nu +  \gamma_\nu p_\mu )
 \pslash \ggcf \frac{1}{2} (-1 + \alpha ) \\
 \lefteqn{ {\overline \Delta}^{\tau_3^{(6)}} (p,m) = 
 {\rm i} \frac{1}{2} ( \gamma_\mu p_\nu +  \gamma_\nu p_\mu ) }
 \nonumber \\
 && + {\rm i} \frac{1}{2} ( \gamma_\mu p_\nu +  \gamma_\nu p_\mu )
 \ggcf ( \ln a^2 p^2 -14.27168 - \frac{1}{2} \alpha + 0.19740 \csw ) 
 \nonumber \\ &&
 + \frac{m}{p^2}  \frac{1}{2} \pslash 
 ( \gamma_\mu p_\nu +  \gamma_\nu p_\mu )
 \ggcf \frac{1}{2} (-1 + \alpha ) .
 \end{eqnarray} 
In the $\MSB$ scheme we have 
 \begin{eqnarray} 
 \lefteqn{ \Lambda_\msb (p, m_\msb ) =  
  {\rm i} \frac{1}{2} ( \gamma_\mu p_\nu +  \gamma_\nu p_\mu ) } 
 \nonumber \\ && 
 + {\rm i} \frac{1}{2} ( \gamma_\mu p_\nu +  \gamma_\nu p_\mu )
 \ggcf \Big[ 
 (\frac{8}{3} -\alpha ) \ln \frac{p^2}{\mu^2} - \frac{31}{9} \Big] 
 \nonumber \\  
 &&+ {\rm i} \frac{ p_\mu p_\nu}{p^2} \pslash \ggcf 
 \Big[ - \frac{1}{3} - \alpha \Big] 
  - m \frac{ p_\mu p_\nu}{p^2} \ggcf 4 . 
 \end{eqnarray}

 \subsection{First unpolarised moment, diagonal, traceless\label{sectunpole}}

 In the  $\tau_1^{(3)}$ (i.e. diagonal, traceless) 
 representation of the hypercubic group is 
  \begin{eqnarray}
 \calO_{\mu\nu}^{\tau_1^{(3)}} & = &
 \left(1+ a\, m\, c_0 \right) \frac{1}{2} \; \psib \frac{1}{2} 
 \left(\gamma_\mu \Dlr_\mu - \gamma_\nu \Dlr_\nu \right) \psi
 \nonumber \\ &&
  + \frac{1}{8}\,a\,{\rm i} c_1  \sum_\lambda \psib
 \frac{1}{2} \left(
 \sigma_{\mu\lambda}\left[\Dlr_\mu,\Dlr_\lambda\right]
 -\sigma_{\nu\lambda}\left[\Dlr_\nu,\Dlr_\lambda\right]
  \right) \psi \nonumber \\
 & & -\frac{1}{8} \,a \,c_2
  \psib\left(\Dlr_\mu \Dlr_\mu - \Dlr_\nu \Dlr_\nu\right)\psi \, . 
 \end{eqnarray}
 In this expression, repeated $\mu$ and $\nu$ indices are not summed over: 
 \begin{eqnarray}
 \lefteqn{ \Lambda^{\tau_1^{(3)}}(p,m;0,c_1,c_2)  
  = {\rm i} \frac{1}{2} ( \gamma_\mu p_\mu  -\gamma_\nu p_\nu ) 
  + c_2 a \frac{1}{2} ( p_\mu p_\mu - p_\nu p_\nu ) }
 \nonumber \\
  & + & {\rm i} \frac{1}{2} ( \gamma_\mu p_\mu  -\gamma_\nu p_\nu ) 
 \ggcf \Big[ 
 (\frac{8}{3} -\alpha ) \ln a^2 p^2 
 -17.52702 + 3.79201 \alpha 
 \nonumber \\ && 
 - 1.72093 \csw + 0.35754 \csw^2 
 + c_1  (-4.27417 + 1.08793 \csw)
 \nonumber \\ && 
 + c_2 ( -6.67330 + 4.53710 \csw + 0.44621 \csw^2 ) \Big] 
 \nonumber \\  
 & + & {\rm i} \frac{1}{2} \frac{(p_\mu p_\mu - p_\nu p_\nu)}{p^2}
  \pslash \ggcf 
 \Big[ - \frac{1}{3} - \alpha \Big] 
  - m \frac{1}{2} \frac{(p_\mu p_\mu - p_\nu p_\nu)}{p^2} \ggcf 4 
 \nonumber \\  
 & + & a \frac{1}{2} (p_\mu p_\mu - p_\nu p_\nu) \ggcf \Big[ 
 (3 -\alpha -\frac{4}{3} \csw ) \ln a^2 p^2 
 \nonumber \\ && 
 -4.58870 + 1.43576 \, \alpha + 1.14260 \csw - 0.07987 \csw^2
 \nonumber \\ && 
 + c_1 \left( - \frac{1}{3} \ln a^2 p^2 + 1.88483 + 1.45305 \csw \right)
 \nonumber \\ && 
 + c_2 \left( (\frac{4}{3} -\alpha )  \ln a^2 p^2 
 -35.68903 + 4.29200 \, \alpha  \right. \nonumber \\ && \left. 
 + 0.97769 \csw - 0.04918 \csw^2
 \right) \Big] 
 \nonumber \\  
  & + & {\rm i} a m \frac{1}{2} ( \gamma_\mu p_\mu  -\gamma_\nu p_\nu ) 
 \ggcf \Big[
 \left( -\frac{13}{6} + \alpha + \frac{1}{2} \csw \right) \ln a^2 p^2
 \nonumber \\ && 
 + 1.00108 - 2.93576 \alpha + 2.62151 \csw -0.74162 \csw^2 
 \nonumber \\ &&  
  + c_1 \left(  \ln a^2 p^2 -2.06261 - 0.62935 \csw \right) 
 \nonumber \\ &&  
 + c_2 \left( \frac{11}{3} \ln a^2 p^2 
 + 0.38293 - \alpha - 4.95277 \csw - 0.20168 \csw^2
 \right) 
 \Big] \nonumber \\   
 & + & a m\,  {\rm i} \frac{1}{2} \,\frac{(p_\mu p_\mu - p_\nu p_\nu)}{p^2}
 \pslash \ggcf
 \Big[ \frac{7}{3} + 2 \alpha + \csw 
 + c_2 ( \frac{2}{3} - 2 \alpha ) \Big] .
 \label{unfirste}
 \end{eqnarray}

 The contact Green's function is given by 
 \begin{equation}
 C^\calO(p,m) = \Delta^\calO (p,m) S(p,m)        
 + S(p,m) {\overline \Delta}^\calO (p,m) \,, 
 \end{equation} 
 where 
 \begin{eqnarray}
 \lefteqn{ \Delta^{\tau_1^{(3)}} 
 = {\rm i} \frac{1}{2} ( \gamma_\mu p_\mu -  \gamma_\nu p_\nu ) }
 \nonumber \\
 && + {\rm i} \frac{1}{2} ( \gamma_\mu p_\mu -  \gamma_\nu p_\nu )
 \ggcf ( \ln a^2 p^2 -14.27168 - \frac{1}{2} \alpha + 0.19740 \csw ) 
 \nonumber \\ &&
 + \frac{m}{p^2}  ( \gamma_\mu p_\mu -  \gamma_\nu p_\nu )
 \pslash \ggcf \frac{1}{2} (-1 + \alpha ) \\
 \lefteqn{  
  {\overline \Delta}^{\tau_1^{(3)}} (p,m) =
 {\rm i} \frac{1}{2} ( \gamma_\mu p_\mu -  \gamma_\nu p_\nu ) }
 \nonumber \\
 && + {\rm i} ( \gamma_\mu p_\mu -  \gamma_\nu p_\nu )
 \ggcf ( \ln a^2 p^2 -14.27168 - \frac{1}{2} \alpha + 0.19740 \csw ) 
 \nonumber \\ &&
 + \frac{m}{p^2}  \pslash 
 \frac{1}{2} ( \gamma_\mu p_\mu -  \gamma_\nu p_\nu )
 \ggcf \frac{1}{2} (-1 + \alpha ) . 
 \end{eqnarray} 

In the $\MSB$ scheme we have 
 \begin{eqnarray} 
 \lefteqn{ \Lambda_\msb (p, m_\msb ) = 
  {\rm i} \frac{1}{2} ( \gamma_\mu p_\mu -  \gamma_\nu p_\nu ) }
 \nonumber \\ && 
 + {\rm i} \frac{1}{2} ( \gamma_\mu p_\mu -  \gamma_\nu p_\nu )
 \ggcf \Big[ 
 (\frac{8}{3} -\alpha ) \ln \frac{p^2}{\mu^2}  - \frac{31}{9} \Big] 
 \nonumber \\  
 &&+ {\rm i}\, \frac{1}{2}
 \frac{(p_\mu p_\mu - p_\nu p_\nu)}{p^2} \,\pslash \,\ggcf 
 \Big[ - \frac{1}{3} - \alpha \Big] 
 \nonumber \\  
  &&- m \frac{1}{2} \frac{(p_\mu  p_\mu - p_\nu p_\nu)}{p^2} \ggcf 4 .
 \end{eqnarray}

 \subsection{ First polarised moment, off-diagonal ($\mu \ne \nu$),
 symmetrised \label{sectpold}}

 For the $\tau_4^{(6)}$ representation we use the operator
 \begin{eqnarray} 
  \calO_{\mu\nu}^{\tau_4^{(6)}} & = &  
 \left(1+ a\, m\, c_0 \right) \frac{1}{2} \; 
 \psib \frac{1}{2}\left(\gamma_\mu\gamma_5 \Dlr_\nu 
 + \gamma_\nu\gamma_5 \Dlr_\mu \right) \psi
 \nonumber \\ &&
 - \frac{1}{4}\, {\rm i} a\, c_1\, \psib\sigma_{\mu \nu}  
 \gamma_5 \frac{1}{2}\left( \Dlr_\nu^2 -  \Dlr_\mu^2 \right) \psi
 \nonumber \\   & & 
 -\frac{1}{8}\,a \,{\rm i}\,c_2 \, \sum_{\lambda \ne \mu, \nu} 
 \psib \frac{1}{2}
 \left( \sigma_{\mu\lambda} 
  \gamma_5\left\{\Dlr_\lambda,\Dlr_\nu\right\}
 + \sigma_{\nu\lambda}
  \gamma_5\left\{\Dlr_\lambda,\Dlr_\mu\right\} \right) \psi  \; .
 \end{eqnarray}
 Repeated $\mu$ or $\nu$ indices are not summed over. 
 The amputated Green's function is:
 \begin{eqnarray}
 \lefteqn{ \Lambda^{\tau_4^{(6)}}(p,m;0,c_1,c_2)  = 
 {\rm i} \frac{1}{2} (\gamma_\mu \gamma_5 p_\nu + \gamma_\nu \gamma_5 p_\mu )}
 \nonumber \\
 &+& c_1 a{\rm i}\frac{1}{2} \sigma_{\mu\nu}\gamma_5 (p_\nu p_\nu -p_\mu p_\mu)
   +  c_2 a {\rm i} \frac{1}{2} \sum_{\lambda \neq \mu,\nu}
 (\sigma_{\mu\lambda} \gamma_5 p_\nu p_\lambda 
 +\sigma_{\nu\lambda} \gamma_5 p_\mu p_\lambda )
 \nonumber \\
 &+& {\rm i}\frac{1}{2} (\gamma_\mu \gamma_5 p_\nu +\gamma_\nu \gamma_5 p_\mu )
 \ggcf \Big[ 
 (\frac{8}{3} -\alpha ) \ln a^2 p^2 
 \nonumber \\ && 
 -19.74374 + 3.79201 \, \alpha 
 +0.88956 \csw -0.49529 \csw^2 
 \nonumber \\ && 
 + c_1\, ( -5.61603 + 4.10778 \, \csw -0.26315 \,\csw^2 )
 \nonumber \\ && 
 + c_2 \, ( -8.29791 + 4.21724 \, \csw -0.49384 \,\csw^2 ) \Big] 
 \nonumber \\  
 & + & {\rm i} \frac{ p_\mu p_\nu}{p^2} \pslash \gamma_5 \ggcf 
 \Big[ - \frac{1}{3} - \alpha \Big] 
 \nonumber \\  
 & + & a {\rm i} \frac{1}{2} 
 \sigma_{\mu\nu} \gamma_5 (p_\nu p_\nu -p_\mu p_\mu) \ggcf \Big[ 
 (\frac{4}{3} -\alpha +\frac{1}{3} \csw ) \ln a^2 p^2 
 \nonumber \\ && 
  -4.96628 +2.43576 \, \alpha +0.86287 \,\csw + 0.06741 \,\csw^2
 \nonumber \\ && 
 + c_1 \big( (\frac{8}{3} - \alpha) \ln a^2 p^2 - 34.73952
  +3.62534 \, \alpha 
 \nonumber \\ && 
 - 0.23227 \csw + 0.01777 \csw^2 \big)
 \nonumber \\ && 
 + c_2 \left( - \frac{5}{3} \ln a^2 p^2 
 +2.39179 + \frac{2}{3} \alpha 
 +0.14576 \csw -0.00452 \csw^2 \right) \Big] 
 \nonumber \\  
 & + & a {\rm i} \frac{1}{2} \sum_{\lambda \neq \mu,\nu}
 (\sigma_{\mu\lambda} \gamma_5 p_\nu p_\lambda 
 +\sigma_{\nu\lambda} \gamma_5 p_\mu p_\lambda ) \ggcf \Big[ 
 (\frac{4}{3} -\alpha +\frac{1}{3} \csw ) \ln a^2 p^2 
 \nonumber \\ && 
  -3.13860 +2.43576 \alpha
  +0.28852 \csw -0.00271 \csw^2
 \nonumber \\ && 
 + c_1 \left( -\frac{5}{3} \ln a^2 p^2 
 + 1.53155 + \frac{2}{3} \alpha - 0.30307 \csw + 0.01158 \csw^2 \right)
 \nonumber \\ && 
 + c_2 \big( (\frac{8}{3} - \alpha) \ln a^2 p^2 
 -32.34394 + 3.62534 \alpha 
 \nonumber \\ && 
 -0.43312 \csw -0.03719 \csw^2 \big) \Big] 
 \nonumber \\  
 &+&  {\rm i} \frac{m}{p^2} \frac{1}{2} \sum_\lambda
 \left( \sigma_{\mu \lambda} \gamma_5 p_\nu p_\lambda 
 + \sigma_{\nu \lambda} \gamma_5 p_\mu p_\lambda \right) 
 \ggcf \left[ -2 + 2 \alpha \right] 
 \nonumber \\  
 &+& {\rm i} a m \frac{1}{2} ( \gamma_\mu \gamma_5 p_\nu 
 +  \gamma_\nu \gamma_5 p_\mu )
 \ggcf \Big[
 \left( -\frac{1}{2} + \alpha - \frac{7}{6} \csw \right) \ln a^2 p^2
 \nonumber \\ &&  
 + 4.77328 - 3.93576 \alpha
 -1.36644 \csw +1.05432 \csw^2 
 \nonumber \\ &&  
 + c_1 \left( \frac{7}{3} \ln a^2 p^2 - 2.30755 -\frac{1}{3} \alpha 
   - 2.28022 \csw + 0.07834 \csw^2 \right) 
 \nonumber \\ &&  
 + c_2 \left( \frac{7}{3} \ln a^2 p^2  + 0.38901 
 - \frac{5}{3} \alpha - 2.29812 \csw +0.21844 \csw^2 \right) 
 \Big] \nonumber \\   
 & + & a m  {\rm i} \frac{ p_\mu p_\nu}{p^2} \pslash \gamma_5 \ggcf
 \Big[-1 +2 \alpha + \frac{1}{3} \csw - \frac{2}{3} (c_1 +c_2) \Big]  
 \nonumber \\
 & + & a {\rm i} \frac{1}{2} \frac{m}{p^2} 
 \left( \gamma_\mu \gamma_5 p_\nu^3 +  \gamma_\nu \gamma_5 p_\mu^3 \right)
 \ggcf (c_1 - c_2) \left[ \frac{8}{3} - \frac{4}{3} \alpha \right] \;. 
 \label{pfirstd}
 \end{eqnarray}
 The contact Green's function is given by 
 \begin{equation}
 C^\calO(p,m) = \Delta^\calO (p,m) S(p,m)        
 + S(p,m) {\overline \Delta}^\calO (p,m) \,, 
 \end{equation} 
 where 
 \begin{eqnarray}
 \lefteqn{ \Delta^{\tau_4^{(6)}} (p,m) =
 {\rm i} \frac{1}{2} ( \gamma_\mu \gamma_5 p_\nu 
 +  \gamma_\nu \gamma_5 p_\mu ) }
 \nonumber \\
 && + {\rm i} \frac{1}{2} ( \gamma_\mu \gamma_5 p_\nu 
 +  \gamma_\nu \gamma_5 p_\mu )
 \ggcf ( \ln a^2 p^2 -14.27168 - \frac{1}{2} \alpha + 0.19740 \csw ) 
 \nonumber \\ &&
 + \frac{m}{p^2}  \frac{1}{2} ( \gamma_\mu \gamma_5 p_\nu
   +  \gamma_\nu \gamma_5 p_\mu )
 \pslash \ggcf \frac{1}{2} (-1 + \alpha ) \,, \\
 \lefteqn{ {\overline \Delta}^{\tau_4^{(6)}} (p,m) = 
 {\rm i} \frac{1}{2} ( \gamma_\mu \gamma_5 p_\nu 
 +  \gamma_\nu \gamma_5 p_\mu ) }
 \nonumber \\
 && + {\rm i} \frac{1}{2} ( \gamma_\mu \gamma_5 p_\nu 
 +  \gamma_\nu \gamma_5 p_\mu )
 \ggcf ( \ln a^2 p^2 -14.27168 - \frac{1}{2} \alpha + 0.19740 \csw ) 
 \nonumber \\ &&
 + \frac{m}{p^2}  \frac{1}{2} \pslash ( \gamma_\mu \gamma_5 p_\nu
   +  \gamma_\nu \gamma_5 p_\mu )
  \ggcf \frac{1}{2} (-1 + \alpha ) \,. 
 \end{eqnarray} 

 In the $\MSB$ scheme we have 
 \begin{eqnarray} 
 \lefteqn{ \Lambda_\msb (p, m_\msb ) =  
 {\rm i} \frac{1}{2} (\gamma_\mu \gamma_5 p_\nu +\gamma_\nu \gamma_5 p_\mu)}
 \nonumber \\ && 
 + {\rm i} \frac{1}{2} ( \gamma_\mu \gamma_5 p_\nu +\gamma_\nu \gamma_5 p_\mu )
 \ggcf \Big[ 
 (\frac{8}{3} -\alpha ) \ln \frac{p^2}{\mu^2} - \frac{31}{9} \Big] 
 \nonumber \\  
 &&+ {\rm i} \frac{ p_\mu p_\nu}{p^2} \pslash \gamma_5 \ggcf 
 \Big[ - \frac{1}{3} - \alpha \Big]  
 \nonumber \\  &&
  +   {\rm i} \frac{1}{2} \frac{m}{p^2} \sum_\lambda
 \left( \sigma_{\mu \lambda} \gamma_5 p_\nu p_\lambda 
 + \sigma_{\nu \lambda} \gamma_5 p_\mu p_\lambda \right) 
 \ggcf \left[ -2 +2 \, \alpha \right] \; . 
 \end{eqnarray} 

 \subsection{ First polarised moment, diagonal, traceless \label{sectpolu}}

 For the $\tau_4^{(3)}$ representation we use the operator
 \begin{eqnarray} 
 \calO_{\mu\nu}^{\tau_4^{(3)}} & = &
 \left(1+ a\, m\, c_0 \right)  \frac{1}{2} \; 
 \psib \frac{1}{2} \left( \gamma_\mu\gamma_5 \Dlr_\mu 
 - \gamma_\nu\gamma_5 \Dlr_\nu \right) \psi
 \nonumber \\ &&
 -\frac{1}{8}\,a \,{\rm i}\,c_1 \, \sum_\lambda \psib
 \frac{1}{2} \left( \sigma_{\mu\lambda}
  \gamma_5\left\{\Dlr_\lambda,\Dlr_\mu\right\}
 -  \sigma_{\nu\lambda}
  \gamma_5\left\{\Dlr_\lambda,\Dlr_\nu\right\} \right)\psi \;. 
 \end{eqnarray} 
 Repeated $\mu$ or $\nu$ indices are not summed over. 
 The operator vertex is:
 \begin{eqnarray}
 \lefteqn{ \Lambda^{\tau_4^{(3)}}(p,m;0,c_1)  = 
 {\rm i} \frac{1}{2} (\gamma_\mu \gamma_5 p_\mu - \gamma_\nu \gamma_5 p_\nu )}
 \nonumber \\
 &+& c_1 a{\rm i}\frac{1}{2} \sum_\lambda
 (\sigma_{\mu\lambda} \gamma_5 p_\mu p_\lambda 
 -\sigma_{\nu\lambda} \gamma_5 p_\nu p_\lambda )
 \nonumber \\
 &+& {\rm i}\frac{1}{2} (\gamma_\mu \gamma_5 p_\mu -\gamma_\nu \gamma_5 p_\nu)
 \ggcf \Big[ 
 (\frac{8}{3} -\alpha ) \ln a^2 p^2 
 \nonumber \\ && 
 -19.92148 + 3.79201 \, \alpha +0.99934 \csw -0.60077 \csw^2 
 \nonumber \\ && 
 + c_1\, ( -15.31376 + 8.54773 \, \csw -0.26036 \,\csw^2 ) \Big]
 \nonumber \\  
 & + & {\rm i} \frac{1}{2} \frac{(p_\mu^2 - p_\nu^2)}{p^2} \pslash \gamma_5 
 \ggcf 
 \Big[ - \frac{1}{3} - \alpha \Big] 
 \nonumber \\  
 & + & a {\rm i} \frac{1}{2} \sum_\lambda
 (\sigma_{\mu\lambda} \gamma_5 p_\mu p_\lambda
 -\sigma_{\nu\lambda} \gamma_5 p_\nu p_\lambda )
 \ggcf \Big[ (\frac{4}{3} -\alpha +\frac{1}{3} \csw ) \ln a^2 p^2 
 \nonumber \\ && 
  -2.72696 +2.43576 \, \alpha
  +0.35052 \,\csw - 0.127617 \,\csw^2
 \nonumber \\ && 
 + c_1 \big( (1 - \alpha) \ln a^2 p^2 - 30.32684
  +4.29201 \, \alpha 
 \nonumber \\ && 
 - 0.64293 \csw - 0.00597 \csw^2 \big) \Big] 
 \nonumber \\  
 &+&  {\rm i} \frac{1}{2} \frac{m}{p^2} \sum_\lambda
 \left( \sigma_{\mu \lambda} \gamma_5 p_\mu p_\lambda 
 - \sigma_{\nu \lambda} \gamma_5 p_\nu p_\lambda \right) 
 \ggcf \left[ -2 + 2 \, \alpha \right] 
 \nonumber \\  
 & + & a m {\rm i} \frac{1}{2} 
 (\gamma_\mu \gamma_5 p_\mu - \gamma_\nu \gamma_5 p_\nu ) \ggcf \Big[ 
 (-\frac{1}{2} +\alpha -\frac{7}{6} \csw ) \ln a^2 p^2 
 \nonumber \\ && 
  +3.94429 -3.93576 \alpha
  -1.41190 \csw +0.90653 \csw^2
 \nonumber \\ && 
 + c_1 \left( \frac{14}{3} \ln a^2 p^2 
 -2.33015 - 2\, \alpha - 4.68331 \csw - 0.03571 \csw^2 \right) \Big]
 \nonumber \\  
 & + & a m  {\rm i} \frac{1}{2} 
 \frac{(p_\mu^2 -p_\nu^2)}{p^2} \pslash \gamma_5 \ggcf
 \Big[-1 +2 \alpha + \frac{1}{3} \csw - \frac{4}{3} c_1  \Big]  \, .
 \end{eqnarray}
 The contact Green's function is given by 
 \begin{equation}
 C^\calO(p,m) = \Delta^\calO (p,m) S(p,m)        
 + S(p,m) {\overline \Delta}^\calO (p,m) \,,  
 \end{equation} 
 where 
 \begin{eqnarray}
 \lefteqn{ \Delta^{\tau_4^{(3)}} (p,m) =
 {\rm i} \frac{1}{2} ( \gamma_\mu \gamma_5 p_\mu 
 -  \gamma_\nu \gamma_5 p_\nu ) }
 \nonumber \\
 && + {\rm i} \frac{1}{2} ( \gamma_\mu \gamma_5 p_\mu 
 -  \gamma_\nu \gamma_5 p_\nu )
 \ggcf ( \ln a^2 p^2 -14.27168 - \frac{1}{2} \alpha + 0.19740 \csw ) 
 \nonumber \\ &&
 + \frac{m}{p^2}  \frac{1}{2} ( \gamma_\mu \gamma_5 p_\mu
   -  \gamma_\nu \gamma_5 p_\nu )
 \pslash \ggcf \frac{1}{2} (-1 + \alpha ) \,, \\
 \lefteqn{ {\overline \Delta}^{\tau_4^{(3)}} (p,m) = 
 {\rm i} \frac{1}{2} ( \gamma_\mu \gamma_5 p_\mu 
 - \gamma_\nu \gamma_5 p_\nu ) }
 \nonumber \\
 && + {\rm i} \frac{1}{2} ( \gamma_\mu \gamma_5 p_\mu 
 - \gamma_\nu \gamma_5 p_\nu )
 \ggcf ( \ln a^2 p^2 -14.27168 - \frac{1}{2} \alpha + 0.19740 \csw ) 
 \nonumber \\ &&
 + \frac{m}{p^2}  \frac{1}{2} \pslash ( \gamma_\mu \gamma_5 p_\mu
   -  \gamma_\nu \gamma_5 p_\nu )
  \ggcf \frac{1}{2} (-1 + \alpha ) \,. 
 \end{eqnarray} 

 In the $\MSB$ scheme we have 
 \begin{eqnarray} 
 \lefteqn{ \Lambda_\msb (p, m_\msb ) =  
  {\rm i} \frac{1}{2} (\gamma_\mu \gamma_5 p_\mu -\gamma_\nu \gamma_5 p_\nu)
 } \nonumber \\ && 
 + {\rm i} \frac{1}{2} ( \gamma_\mu \gamma_5 p_\mu -\gamma_\nu \gamma_5 p_\nu )
 \ggcf \Big[ 
 (\frac{8}{3} -\alpha ) \ln \frac{p^2}{\mu^2} - \frac{31}{9} \Big] 
 \nonumber \\  
 &&+ {\rm i} \frac{1}{2}
 \frac{(p_\mu^2- p_\nu^2)}{p^2} \pslash \gamma_5 \ggcf 
 \Big[ - \frac{1}{3} - \alpha \Big]  
 \nonumber \\  &&
  +   {\rm i} \frac{1}{2} \frac{m}{p^2} \sum_\lambda
 \left( \sigma_{\mu \lambda} \gamma_5 p_\mu p_\lambda 
 - \sigma_{\nu \lambda} \gamma_5 p_\nu p_\lambda \right) 
 \ggcf \left[  -2 + 2 \, \alpha \right] \; . 
 \end{eqnarray}

 \end{document}